\documentclass[useAMS,usenatbib]{mn2e}
\usepackage{graphicx}
\usepackage{txfonts}
\usepackage{url}
\usepackage[dvips,usenames]{color}

\newcommand{\sub}[1]{\mbox{$_{\rm #1}$}}

\newcommand{\Teff}{\mbox{$T\sub{eff}$}}

\newcommand{\beq}{\begin{equation}}
\newcommand{\eeq}{\end{equation}}
\newcommand{\beqa}{\begin{eqnarray}}
\newcommand{\eeqa}{\end{eqnarray}}
\newcommand{\benu}{\begin{enumerate}}
\newcommand{\eenu}{\end{enumerate}}
\newcommand{\bite}{\begin{itemize}}
\newcommand{\eite}{\end{itemize}}
\newcommand{\bdes}{\begin{description}}
\newcommand{\edes}{\end{description}}

\title[Dust in TP-AGB Stars]{Evolution of thermally pulsing asymptotic giant branch
stars III. Dust production at supersolar metallicities}

\author[Nanni et al.]{Ambra Nanni$^1$, Alessandro Bressan$^1$, Paola Marigo$^2$,
  L\'eo Girardi$^{3}$
  \\
  $^1$ SISSA, via Bonomea 265, I-34136 Trieste, Italy \\
  $^2$ Dipartimento di Fisica e Astronomia Galileo Galilei,
  Universit\`a di Padova, Vicolo dell'Osservatorio 3, I-35122 Padova, Italy \\
  $^3$ Osservatorio Astronomico di Padova, Vicolo dell'Osservatorio 5,
  I-35122 Padova, Italy \\
  Database publicly available at: http://www.sissa.it/ap/research/dustymodels.php
}

\begin{document}

\date{Accepted 2013 December 3.  Received 2013 December 3; in original form 2013 September 30}

\pagerange{\pageref{firstpage}--\pageref{lastpage}} \pubyear{2013}

\maketitle

\label{firstpage}

\begin{abstract}
We extend the formalism presented in our recent calculations of dust ejecta from
the Thermally Pulsing Asymptotic Giant Branch (TP-AGB) phase,
to the case of super-solar metallicity stars.
The TP-AGB evolutionary models are computed with the \textsc{colibri} code.
We adopt our preferred scheme for dust growth.
For M-giants, we neglect chemisputtering by H$_2$ molecules and,
for C-stars  we assume a homogeneous growth scheme which is primarily
controlled by the carbon over oxygen excess.

At super-solar metallicities, dust forms more efficiently
and  silicates tend to condense significantly closer to the photosphere ($r\sim 1.5$~$R_*$) -- and thus at higher temperatures
and densities -- than at solar and sub-solar metallicities ($r\sim 2$-$3$~$R_*$).

In such conditions, the hypothesis of thermal decoupling between gas and dust
becomes questionable, while dust heating due to collisions plays an important role.
The heating mechanism delays dust condensation to slightly outer regions
in the circumstellar envelope.
We find that the same mechanism is not significant at solar and sub-solar metallicities.

The main dust products at super-solar metallicities are silicates.
We calculate the total dust ejecta and dust-to-gas ejecta,
for various  values of the stellar initial masses and initial metallicities $Z=0.04,\,0.06$.

Merging these new calculations with those for lower metallicities it turns out that,
contrary to what often assumed, the total dust-to-gas ejecta of intermediate-mass stars
exhibit only a weak dependence on the initial metal content.

\end{abstract}


\section{Introduction}
The production of dust
during the Thermally Pulsing Asymptotic Giant Branch (TP-AGB) evolution of low- and
intermediate-mass stars has been recently revisited by
\citet{Zhukovska08}, \citet{ventura12, DiCriscienzo_etal13} and by our group
\citep{Nanni13}, following the scheme pioneered by \citet{GS99} and \citet{FG06}.

Making use of the TP-AGB models computed with the new \textsc{colibri}
code by \citep{marigoetal13}, we have
investigated the contribution to the interstellar medium (ISM) dust budget
of TP-AGB stars from low metallicities ($Z=0.001$) to the solar one ($Z=0.02$).

At low metallicities ($Z=0.001$) , our dust ejecta, mainly of carbonaceous type,
turn out to be more than one order of magnitude larger than those computed
by \citet{ventura12} for almost all stellar initial masses.
This is a direct consequence of the larger carbon excess
reached by our carbon star models of low metallicities, as a cumulative result of the
differences in the efficiency of the third dredge-up and hot-bottom burning,
and in the duration of the TP-AGB, between our present reference version\footnote{We recall
that a fine calibration of the TP-AGB phase as a function of stellar mass and metallicity
is currently underway so that new versions of the TP-AGB models computed
with \textsc{colibri} are also planned.}
of the \textsc{colibri}  TP-AGB models and  those of \citet{ventura12}.

Needless to say, a higher dust production at low metallicities could have a far-reaching relevance
in the broader context of galaxy formation and evolution at high redshifts where,
observations of galaxies and quasars
indicate that even very young objects can contain
large dust reservoirs \citep{Lilly99, Eales00, Bertoldi03, Robson04, Beelen06, Dwek11}.
Such large amounts of dust at early epochs, when the dust and chemical enrichment time-scales are
only a fraction of a Gyr, are difficult to explain \citep{Maiolino04, Marchenko06,
Dwek07, Valiante09,Mattsson11, Valiante11, Dwek11, Gall11,
Pipino11a, Pipino11b, Michalowski10, Yamasawa11, Zhukovska2013}.

Unfortunately, it is difficult to check the robustness of these predictions because,
in spite of the huge literature on  carbon stars at such low metallicities,
there is a severe lack of {\it direct} observations.

At solar metallicity, direct estimates of the amount of dust produced by AGB stars and its mineralogy are provided by
mid and far-infrared observations, both in our galaxy and in the nearby ones \citep{Knapp85, Matsuura09, Matsuura12}.
\citet{Knapp85} have found typical dust-to-gas ratios of $\sim$6~$\times$~10$^{-3}$
for oxygen-rich (M) AGB stars, mainly in the form of
amorphous silicates,  and $\sim$10$^{-3}$ for carbon-rich (C) stars,
mainly as amorphous carbon. These ratios indicate that,
at solar metallicity, a large fraction  of the silicon
must condense into dust in the circumstellar envelopes (CSEs) of these stars.
The dust-to-gas ratios of M-giants predicted by \citet{Nanni13} are within the observed range,
but there is a well known problem in reproducing their terminal velocities \citep{Woitke06, Hofner09, Bladh12}.
We have shown that, in order to reproduce
the relation between terminal velocities and mass loss rates observed in Galactic oxygen-rich giants,
one may relax the strong constraint on the condensation temperature of
silicates imposed by the chemisputtering destruction process \citep{Nanni13}.
The higher condensation temperatures obtained by  \citet{Nanni13} by neglecting this destruction process,
are in very good agreement with the laboratory measurements
\citep{Nagahara96, nagaharaetal09}.

In this paper we wish to extend our previous analysis to the case of super metal rich stars.
In our Galaxy, only the Bulge may eventually harbor traces of such stellar populations,
but these could be the major component in the nuclear regions of massive elliptical galaxies
\citep{Bertolaetal95}.
In some of the Virgo cluster early type galaxies, the interstellar medium (ISM) contribution to the Mid Infrared (MIR) emission
 is very low, and Spitzer  spectroscopic observations have revealed the presence of a prominent
9.7~$\mu$m feature, typical of silicates  \citep{Bressanetal06}.
Additional evidence, brought by Spitzer imaging between 4 and 16 ${\mu}$m
and near-infrared data, indicates that this MIR emission is stellar in origin, confirming that
it comes from the dusty envelopes around evolved AGB stars \citep{Clemensetal11}.
It has been suggested that this feature  could be
used as a powerful tool to disentangle the degeneracy between age and metallicity,
an effect that challenges the census of stellar populations in passively evolving early type galaxies \citep{Bressan98}.
While this has been shown to be the case for a metal poor massive star cluster in M81
\citep{Mayyaetal13}, in the case of early type galaxies the results
are much less robust because of the intrinsic difficulties encountered in modeling
the intensity of the MIR emission at high metallicity \citep{Clemensetal09}.
These models are based on empirical scaling relations extrapolated from observations
of AGB stars in the Magellanic Clouds and in the Galaxy, while
direct calculations of the properties of dust in the CSEs, as those of \citet{Nanni13},
should be preferred.

The paper is organized as follows.
In Section~\ref{tpagbmodels} we summarize the main characteristics
of the  TP-AGB tracks at super solar  metallicities.
The basic equations of the wind model presented in \citet{Nanni13} are briefly summarized
in Section~\ref{sec_wind}.
In Section~\ref{sec_growth} we discuss
the equations governing the dust growth with particular focus on the determination
of dust equilibrium temperature.
In Section~\ref{sec_res} we apply the method to the new  models of TP-AGB of high metallicity
while, in Section~\ref{sec_dust_ejecta}, we provide the corresponding
dust and dust-to-gas ejecta for different masses.
Finally, the results are discussed in Section~\ref{sec_discussion}.

\section{The TP-AGB models}\label{tpagbmodels}
Stellar evolution from the pre main sequence up to the first thermal pulse is calculated with the
\texttt{PARSEC} code \citep{Bressanetal12} while the following evolution along the
the whole TP-AGB phase is computed with the \textsc{colibri}
 code \citet{marigoetal13}. The reader should
refer to those papers for all the details on stellar evolution.
\begin{figure*}
\centering
\includegraphics[angle=90,width=0.95\textwidth]{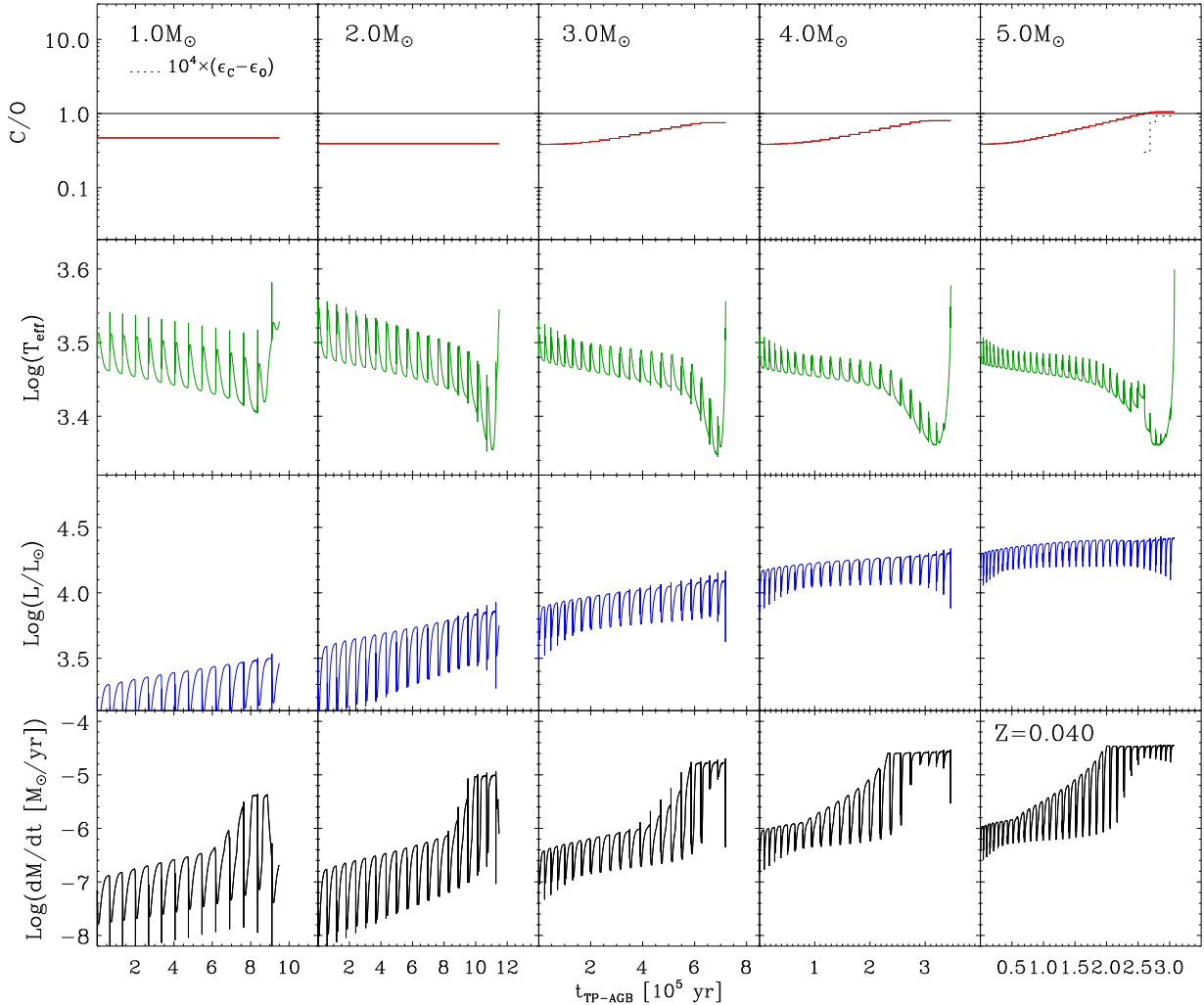}
\caption{Evolution of  surface C/O, carbon excess $\epsilon_{\rm C}$-$\epsilon_{\rm O}$ (only when positive),
 effective temperature, luminosity, and
mass-loss rate during the whole TP-AGB phase
of a few selected models with initial metallicity $Z=0.04$, computed
with the \textsc{colibri} code \citep{marigoetal13}.
These quantities are the key
input stellar parameters for our dust growth model. Time is counted from the first thermal
pulse. Note that effective temperature and luminosity
derive from the solution of the full set
of stellar structure equations across the envelope and the atmosphere,
and not from analytic relations as commonly done in synthetic TP-AGB models.
The value of the C/O ratio is always below unity, but for the late stages
of the model
with initial mass $M=5.0$~M$_{\odot}$.
See the text for more details.}
\label{fig_agbmodz04}
\end{figure*}
\begin{figure*}
\centering
\includegraphics[angle=90,width=0.95\textwidth]{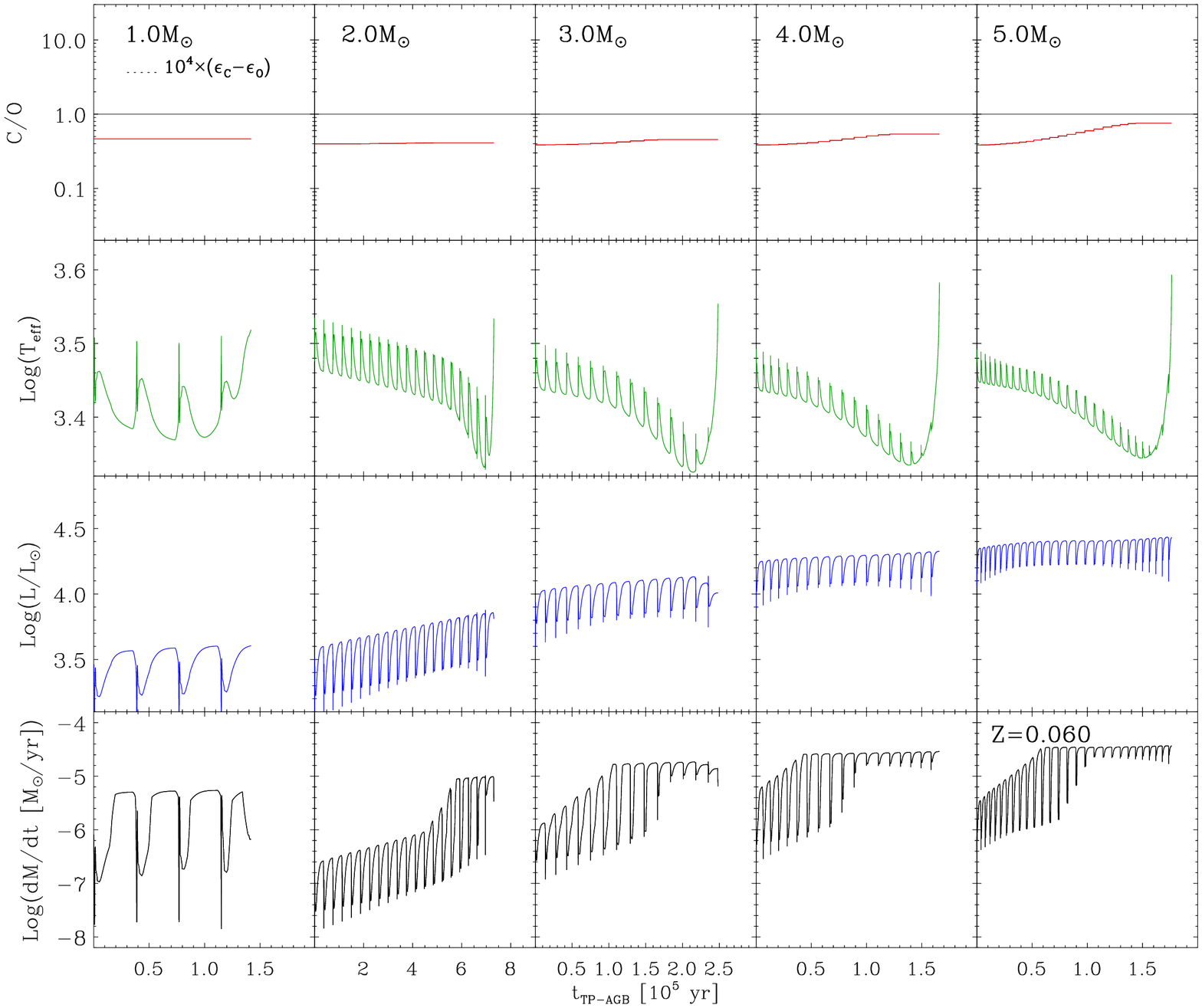}
\caption{The same as in Fig.~\ref{fig_agbmodz04},
 but for initial metallicity $Z=0.06$.}
\label{fig_agbmodz06}
\end{figure*}
We only recall here that, thanks to the incorporation of the
\textsc{\ae sopus} tool \citet{MarigoAringer_09} in the \textsc{colibri}
code our TP-AGB evolution models are the first
ever to include accurate on-the-fly computation of the equation of the equation
of state for $\simeq 300$ atomic and  $\simeq 500$ molecular species,
as well as of the Rosseland mean gas opacities throughout the
convective envelope and the atmosphere.
This new advance guarantees full consistency between the
envelope structure and the surface chemical abundances, and
therefore robustly tracks the impact of third dredge-up
mixing episodes and hot-bottom burning.

In particular, with \textsc{colibri} we are able to follow
in detail the evolution of the surface C/O ratio,
which critically affects molecular chemistry, opacity, and
effective temperature, every time it goes across
the critical values around unity
\citep{Marigo_02, MarigoGirardi_07}.
As a consequence, the C/O ratio plays a major
role in determining the chemical and physical
properties of the dust. As discussed in \citet{Nanni13},
in M-stars (C/O$<1$) the main dust species are
amorphous silicates, quartz (SiO$_2$) and
corundum (Al$_2$O$_3$) \citep{Tielens98, Ossenkopf92} while,
in C-stars (C/O$>1$), the dust produced is
predominantly amorphous carbon and silicon carbide (SiC) \citep{Groenewegen98}.
At super-solar metallicity however, the latter phase is practically absent,
as can be seen from
Figs.~\ref{fig_agbmodz04} and \ref{fig_agbmodz06}.
Only the model of $M=5$~M$_\odot$ with $Z=0.04$ is able to reach
a C/O ratio slightly larger than unity.

The almost complete lack of the C-star phase is the combined result of
the minor efficiency of the third dredge-up and the
shorter lifetimes in TP-AGB models with high initial metallicity.
First, the minimum stellar
mass at which the third dredge-up is expected to take place increases with
the metallicity, while the depth of the third dredge-up becomes
progressively shallower. Second, even TP-AGB models that experience the
third dredge-up may not be able to reach the condition C/O$>1$, because
mass loss by stellar winds prevents them from undergoing the number
of thermal pulses necessary to bring the surface carbon abundance above the
already high initial content of oxygen.
We also underline the fact that at high metallicities,
such those considered in this work $(Z=0.04,\,0.06)$, hot bottom burning
is not active in most of the more massive AGB stars, except for those
with $M\simeq 6$~M$_{\odot}$,
close to upper limit of the mass range covered
by intermediate-mass stars.
As a consequence, these models do not present the overluminosity
effect usually associated to hot-bottom burning, and their luminosity
evolution during the quiescent stages is mainly controlled by the mass
of the H-exhausted core (see third-row panels in Figs.~\ref{fig_agbmodz04}
and \ref{fig_agbmodz06}).

We notice that the effective temperatures of these models
decrease on average, as the metallicity increases, as expected
as a consequence of the higher atmospheric opacity.
Although the surface C/O ratio
remains lower than unity during the whole TP-AGB evolution, the
tracks show a clear and sudden cooling during the late
evolutionary stages (see second-row panels of
Figs.~\ref{fig_agbmodz04} and \ref{fig_agbmodz06}). This happens
as soon as  the TP-AGB models enter the super-wind regime and their
envelope start being dramatically reduced (yet, still not
so much to leave the AGB towards higher $T_{\rm eff}$).
In fact, we know from theory that, at given chemical composition,
the Hayashi lines shift to lower $T_{\rm eff}$ at decreasing stellar mass
(see, e.g., figure 16 of Marigo et al. 2013).

According to our adopted scheme of mass loss
(as detailed in section 6.2 of Marigo et al. 2013), because of
their low effective temperatures,
super-metallicity TP-AGB models quickly
reach the super-wind phase,  which is treated following the
prescriptions of \citet[][their equations 1 and 3]{Vassiliadis93}.
This explains why our predicted mass-loss rates flatten out to nearly constant
values, almost irrespective of both stellar
mass and metallicity (see bottom panels Figs.~\ref{fig_agbmodz04}
and \ref{fig_agbmodz06}).

\section{Wind model}
\label{sec_wind}
The adopted dusty wind model has been thoroughly described in \citet{Nanni13}.

Following FG06,
 the equations below describe a stationary and spherically symmetric outflow
of one-fluid component, assuming that
there is no drift velocity between gas and dust.
Neglecting the contribution of the gas pressure gradient we have
\begin{equation}\label{dvdr}
 v \frac{dv}{dr}=-\frac{G M_*}{r^2}(1-\Gamma)\,,
\end{equation}
where
\begin{equation}\label{gamma}
 \Gamma=\frac{L_*}{4{\rm \pi} c G M_*}\, \kappa
\end{equation}
is the ratio between the radiative and the gravitational accelerations.
The density profile $\rho(r)$ across the wind is determined by the continuity equation:
\begin{equation}\label{rho_pr}
\rho(r)=\frac{\dot{M}}{4 {\rm \pi} r^2 v}\, .
\end{equation}
The temperature structure $T(r)$ is described with
the approximation for a grey and spherically symmetric extended atmosphere
\citep{Lucy71, Lucy76}
\begin{equation}\label{T_pr}
 T(r)^4=\Teff^4 \Big[W(r)+\frac{3}{4}\tau\Big],
\end{equation}
where
\begin{equation}\label{W_r}
 W(r)=\frac{1}{2}\Big[1-\sqrt{1-\Big(\frac{R_*}{r}\Big)^2}\Big],\,
\end{equation}
represents the dilution factor, $R_{*}$ is the photospheric radius,
and $\tau$ is the optical depth that
obeys the differential equation
\begin{equation}\label{dtaudr}
 \frac{d \tau}{d r}=-\rho \kappa \frac{R_{*}^{2}}{r^2},
\end{equation}
with the boundary condition
\begin{equation}\label{taufin}
 \lim_{r\rightarrow\infty}\tau=0.
\end{equation}
The opacity $\kappa$
is given by the sum of the gas contribution and of all the dust species
\begin{equation}\label{kh}
 \kappa= \kappa_{\rm gas}+\sum_i f_i \kappa_{i} \quad [cm^2g^{-1}]
\end{equation}
where $\kappa_{\rm gas}=10^{-8}\rho^{2/3} T^3$ \citep{Bell94}, $f_i$ is the degree of condensation
of the key-element\footnote{Following FG06, the key-element of a given
dust species
denotes the least abundant element among those involved in its formation;
e.g., Si is the key-element of the silicate compounds in O-rich stars.}
 into a certain dust species $i$ and $\kappa_i$ is its
opacity,
computed assuming the complete condensation of the key-element initially available
in the gas phase.
The degree of condensation is usually written as,
\begin{equation}\label{dcond}
 f_i=n_{k,i}\frac{4 {\rm \pi} (a_i^3-a_0^3)\rho_{d,i}}{3 m_{d,i} \epsilon_{k,i}} \epsilon_s,
\end{equation}
where $n_{k,i}$ is the number of atoms of the key-element present
in one monomer of the dust species $i$;
$m_{d,i}$ is the mass of the monomer;
$a_i$ denotes the grain size (radius) and $a_0$ the initial grain size;
$\rho_{d,i}$ is the dust density of the grain;
$\epsilon_{k,i}$, $\epsilon_s$
are the number densities of the key-element,
and of the initial number of dust grains (seed nuclei)
normalized to the number density of hydrogen $N_{H}$, respectively.
Dust opacities are computed from the optical properties of grains, as in \citet{Nanni13}.
For M-stars, the ($n$, $k$) data of olivine and pyroxene are taken
from \citet{Ossenkopf92},  and for corundum we refer to \citet{Begemann97}.
For C-stars the ($n$, $k$) data of amorphous carbon are
from \citet{Hanner88},  while we use \citet{Pegourie88} for silicon carbide.
Iron opacity is derived from \citet{Leksina67}.
Once dust is formed,  its opacity increases and
the contribution of the gas becomes negligible. The wind
accelerates if the opacity becomes large
enough so that $\Gamma>1$.

\subsection{Accretion of dust grains}
\label{sec_grain}
The growth of dust grains
is determined by the balance between the rate of effective collisions
of molecules on the grain surface, $J^{\rm gr}_i$, and its decomposition rate, $J^{\rm dec}_i$.
The differential equation describing the dust growth is usually expressed in terms
of the variation of the
grain radius, $a_i$,
\begin{equation}\label{dadt}
  \frac{da_i}{dt}=V_{0,i}(J^{\rm gr}_i-J^{\rm dec}_i),
 \end{equation}

The growth rate for each dust species, $i$, is defined as the minimum between the
rates of effective collisions on the grain surface
of the gas molecules involved in the formation of the dust monomer through the formation
reactions.
The molecular species determining such a rate  is named ``rate-determining species''.
\begin{equation}\label{growth}
  J^{\rm gr}_i=\min\Big [s_i\frac{\alpha_i n_{j, g} v_{th,j}(T_{\rm gas})}{s_j}\Big ],
\end{equation}
where $n_{j, g}$ is the number density of the molecule $j$ in
the gas phase, $T_{\rm gas}$ is the gas temperature given by Eq.~(\ref{T_pr}), $v_{th, j}(T_{\rm gas})$
the corresponding thermal velocity, $s_j$ its stoichiometric
coefficient in the dust formation reaction, $s_i$ the stoichiometric coefficient of the monomer of the dust species $i$ and $\alpha_i$ is its sticking coefficient.

\section{Dust growth}
\label{sec_growth}
Dust formation in the expanding CSE of an AGB star  can be modeled
as a two-step process. Initially, small stable refractory aggregates, seed nuclei,
are supposed to form from the molecules in the gas phase (nucleation process) and then,
as the temperature decreases below a certain
critical value identified as the dust condensation temperature $T_{\rm cond}$,
accretion on the seed surface occurs by addition
of other molecules.
Following a commonly accepted scenario,  the first to form are
the most refractory aggregates, and then the process
proceeds by heterogeneous accretion \citep{Gail86, Jeong03}.
Since the details of the nucleation process are still poorly understood \citep{Goumans12}
it is common practice to express the abundance of seed nuclei as a
tunable parameter , $\epsilon_s$.
The abundance of seeds usually adopted for solar metallicity, $\epsilon_s=10^{-13}$,
appears to be consistent with detailed nucleation computations by \citet{Jeong03},
as well as in very good agreement with the value inferred by
\citet{Knapp85} for a sample of Galactic M-giants.
Adopting $\epsilon_s=10^{-13}$, \citet{Nanni13} obtained
grains with typical sizes of $\sim 0.15$~$\mu$m, in agreement with the observations.
Similar considerations and  results hold also for amorphous carbon.
In principle, the abundance of seeds should scale with the abundance of the key-element
at varying metallicities.
However, due to the uncertainties affecting the determination of the chemistry of the first
refractory compounds \citep{Jeong03}, we simply scale the seed number with the total metallicity:
\begin{equation}
 \epsilon_{s,M}=\epsilon_s \Big(\frac{Z}{Z_{\rm ISM}}\Big)\,,
\label{nseeds_M}
\end{equation}

where Z$_{\rm ISM}=0.017$ is the assumed local metallicity of the ISM.

\subsection{The dust condensation temperature $T_{\rm cond}$}
\label{ssec_tdust}
As thoroughly discussed in the literature,
the dust condensation temperature, $T_{\rm cond}$, is a critical parameter
that affects the dust condensation radius and consequently the wind dynamics \citep{Hofner_size08, Hofner09, Bladh12}.

In \citet{Nanni13} the condensation radius is defined as the distance from the photosphere
at which the condition $J^{\rm gr}_i=J^{\rm dec}_i$ is satisfied and
 $T_{\rm cond}$ is the equilibrium temperature
that the dust would have at the condensation radius.

The equilibrium temperature is
computed from the balance between the energy absorbed and re-emitted by the dust grains.
A remarkable difference with respect to \citet{Nanni13} is that
here we have
also added the heating rate due to collisions with H$_2$ molecules.
In other works (\citet{GS99} and FG06) this term has been neglected
because, at the low condensation temperatures obtained for silicates and carbon ($T_{\rm gas}\leq{1100K}$),
this term turns out to be negligible. However, in \citet{Nanni13} and in this paper as well,
the condensation temperature for silicates can reach values in excess of $T_{\rm dust}\sim{1400}$~K.
Dust may thus form significantly closer to the photosphere ($r\sim 2$-$3$~$R_*$) than in GS99 and FG06 ($r\sim 5$-$7$~$R_*$),
and, especially at the high metallicities considered in this paper, it is important to
check whether the collisions term could be important. At this metallicities, in fact,
the number density of the particles in the gas phase, and therefore $J^{\rm gr}_i$, are
larger than lower metallicity cases. As a consequence, dust condenses closer to the photosphere, because
the condition $J^{\rm gr}_i=J^{\rm dec}_i$ is reached at smaller radii. At this radii the gas density and
temperature are also larger, and dust and gas might be coupled.

Assuming that at each collision a dust grain gains an energy amount equal to  $3/2 k_B (T_{\rm gas}-T_{\rm dust})$ \citep{Lucy76},
dust heating by gas molecules (in erg~s$^{-1}$) can be expressed as
\begin{equation}\label{collision_rate}
H_{\rm collision}= 4{\rm \pi}{a^2}  3/2 k_B (T_{\rm gas}-T_{\rm dust}) N_{H_2} v_{th}.
\end{equation}
Correspondingly, the energy balance equation used by \citet{Nanni13} for deriving the dust condensation temperature is modified as
\begin{equation}\label{energy_balance}
\sigma~T_{\rm dust}^4 Q_{\rm abs, P}(a, T_{\rm dust})=\sigma~\Teff^4 Q_{\rm abs, P}(a, \Teff) W(r) + \frac{H_{\rm collision}}{4{\rm \pi}{a^2}},
\end{equation}
where $Q_{\rm abs, P}$ is the Planck averaged absorption coefficient expressed explicitly as
a function of the temperature and grain size, $W(r)$ is the dilution factor
 defined in Eq.~(\ref{W_r}) and $N_{\rm H_2}$ is the number density of H$_2$ molecules that is approximately equal to $N_H/2$.
 We stress that the effect of heating by collisions
As discussed in the following, we find that the heating term due to collisions
become important at the highest mass-loss rates, when dust tends to condense close to the photosphere.

\subsubsection{M-star models}
\label{mstarmodels}
For M-giants the main destruction process of dust grains is supposed to be sublimation
caused by heating of the grains due to absorption of stellar radiation
and by collisions with gas particles.
Chemisputtering by H$_2$ molecules, considered to be fully efficient
by \citet{GS99}, is instead neglected. Therefore the decomposition rate is $J^{\rm dec}_i=J^{\rm sub}_i$
This quantity is determined by
considering that
it must equal the growth rate in chemical equilibrium conditions and that,
the rate so determined
also holds outside equilibrium.
We thus obtain from Eq.~(\ref{growth}), after eliminating $n_{j, g}$ with the partial pressure and the temperature,
\begin{equation}
\label{dust_subl}
J^{\rm sub}_i= \alpha_i v_{th, j}(T_{\rm dust}) \frac{p(T_{\rm dust})}{k_B T_{\rm dust}},
\end{equation}
where $T_{\rm dust}$ is the dust equilibrium temperature
determined with Eq.~(\ref{energy_balance}), $v_{th, j}(T_{\rm dust})$
is the thermal velocity of the molecule ejected from the grain surface and $k_B$ is the Boltzmann constant.
The quantity  $p(T_{\rm dust})$ is the saturated vapour pressure at the dust temperature,
that can be expressed with the Clausius-Clapeyron equation
\begin{equation}
\label{ptd}
\log p(T_{\rm dust})=-\frac{c_1}{T_{\rm dust}} + c_2
\end{equation}
where $c_1$ and $c_2$ are sublimation constants, characteristic of the species
under consideration.
The constant $c_1$ contains the latent heat of sublimation of the dust species
and the constant $c_2$ is slightly dependent on the temperature.
The quantities $c_1$ and $c_2$ may be obtained either directly from
thermodynamical data \citep{Duschl96},
or by fitting with Eq.~(\ref{ptd}) the results
of sublimation experiments \citep{Kimura02,Kobayashi09,Kobayashi11}.
We notice that the sublimation rate depends on the
dust equilibrium temperature $T_{\rm dust}$ which, in our model,
can be defined only near the condensation point (Section~\ref{ssec_tdust}).
Thus, instead of integrating the full Eq.~(\ref{dadt}),
we first determine the condensation point within the CSE by comparing the
growth rate with the maximum sublimation rate,
i.e. the sublimation rate obtained by setting $\alpha_i=1$ in Eq.~(\ref{dust_subl}).
This point is defined by the condition $J^{\rm gr}_i = J^{\rm dec}_{i, {\rm max} }$ and provides also
the condensation temperature.
Beyond this point,  we assume that the sublimation rate is negligible
in the right-hand side of Eq.~(\ref{dadt}).
In this way, the condensation temperature
depends both on the dust species which determines $J^{\rm dec}_{i, {\rm max}}$ and
on the physical conditions of the CSE, which determine $J^{\rm gr}_i$.
Since the real sublimation rate is only $\alpha_i$ times the maximum sublimation rate, the above condition
implies that condensation begins when
the growth rate is $\sim$ $1/\alpha_i$ ($\sim$10 for silicates) the
real sublimation rate. The corresponding super-saturation ratio
is also $\sim$1/$\alpha_i$.
Therefore, differently from other models found in the literature,
silicate condensation temperature is not assumed a priori, but is derived
as previously described.
We notice that our choice of retaining only the growth term
in Eq.~(\ref{dadt}) does not affect the accuracy of the results,
since, beyond the condensation point,
the sublimation rate decreases almost exponentially with the
temperature.

\subsubsection{C-star models}
\label{cstarmodels}
For amorphous carbon, we consider the
homogeneous accretion that
initially proceeds through complex reactions of C$_2$H$_2$ addition,
forming  isolated chains that subsequently coalesce into larger cores.
Further growth of carbon mantles on these initial seeds can continue
through vapor condensation \citep{GS99}.
According to \citet{Cherchneff92}, the chain of C$_2$H$_2$ addition reactions
have a bottleneck in the formation of the benzene because it becomes effective
only when the {\it gas}  temperature is below $T_{\rm gas}$=1100~K.
Therefore,  while the sublimation temperature of solid carbon
can exceed $\sim$1600~K, its growth should be inhibited above $T_{\rm gas}$=1100~K.
Thus, following FG06, we do not consider
any destruction reaction in the case of amorphous carbon,
but we assume that it can grow only when $T_{\rm gas}\leq$1100~K.

\subsection{Method of solution and initial conditions}
The system of differential equations describing the dust evolution is given by Eqs.~(\ref{dvdr}), (\ref{dtaudr}) and
an equation like (\ref{dadt}) for each dust species.
In M-stars we consider the evolution of
corundum (Al$_2$O$_3$), quartz (SiO$_2$), iron,
olivine (Mg$_{2x}$Fe$_{2(1-x)}$SiO$_4$)
and pyroxene (Mg$_{x}$Fe$_{1-x}$SiO$_3$).
The quantity $x$, the fractional abundance of Mg molecules with respect to
the sum of Mg and Fe molecules and it ranges from 0 to 1.
In C-stars we consider amorphous carbon, silicon carbide (SiC) and iron.
The independent variable of the system is the radial distance $r$, whereas, the dependent variables are the velocity, $v$, the optical depth,
$\tau$, the grain size for each dust species, $a_i$ and $x$.
The quantities that determine the wind structure are the gas temperature and density profiles, $T(r)$ and $\rho(r)$,
the total opacity, $\kappa$, the dust condensation fraction, $f$, and $\Gamma$.

\begin{figure}
\centering
\includegraphics[angle=0,width=0.45\textwidth]{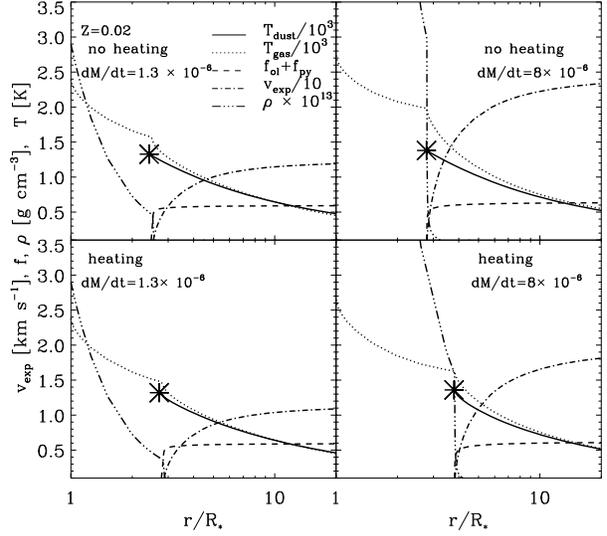}
\caption{Effects of including the collisional heating term in the energy balance equation (Eq. \ref{energy_balance}),
in the structure of the CSE. We show a model with initial mass $M=1.5$~M$_\odot$ and metallicity $Z=0.02$, and
for intermediate (left panels) and high (right panels) mass-loss rates
as indicated, respectively.  In the upper panels the heating term is neglected, as in \citet{Nanni13}
while, in the lower panels, the term is included. In each panel we show the dust equilibrium temperature (solid line),
the gas temperature (dotted line), the condensation fraction of silicates (dashed line), the expansion
velocity (dot-dashed line) and the gas density (triple-dot - dashed line). The asterisk marks the condensation temperature.}
\label{CSE_Z0.02}
\end{figure}
\begin{figure}
\centering
\includegraphics[angle=90,width=0.45\textwidth]{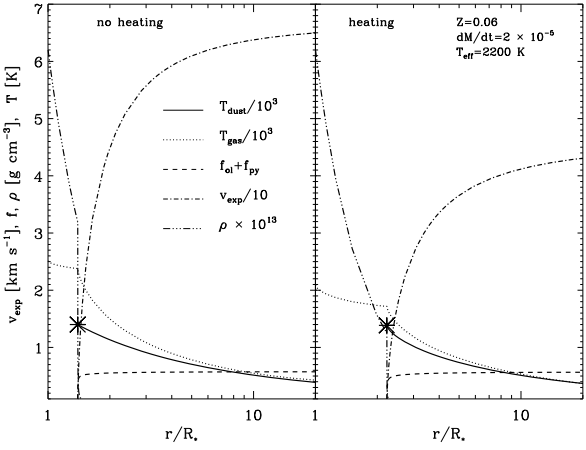}
\caption{As in Fig.~\ref{CSE_Z0.02}, but for a model of  initial mass $M=4$~M$_\odot$,
metallicity $Z=0.06$ and high  mass-loss rate. In the left panel the heating term is neglected
while, in the right  panel, the term is included. Notice that, at this high metallicity,
 neglecting the heating term leads to inconsistent results. See text for more details}
\label{CSE_Z0.06}
\end{figure}

\begin{figure}
\centering
\includegraphics[angle=0,width=0.45\textwidth]{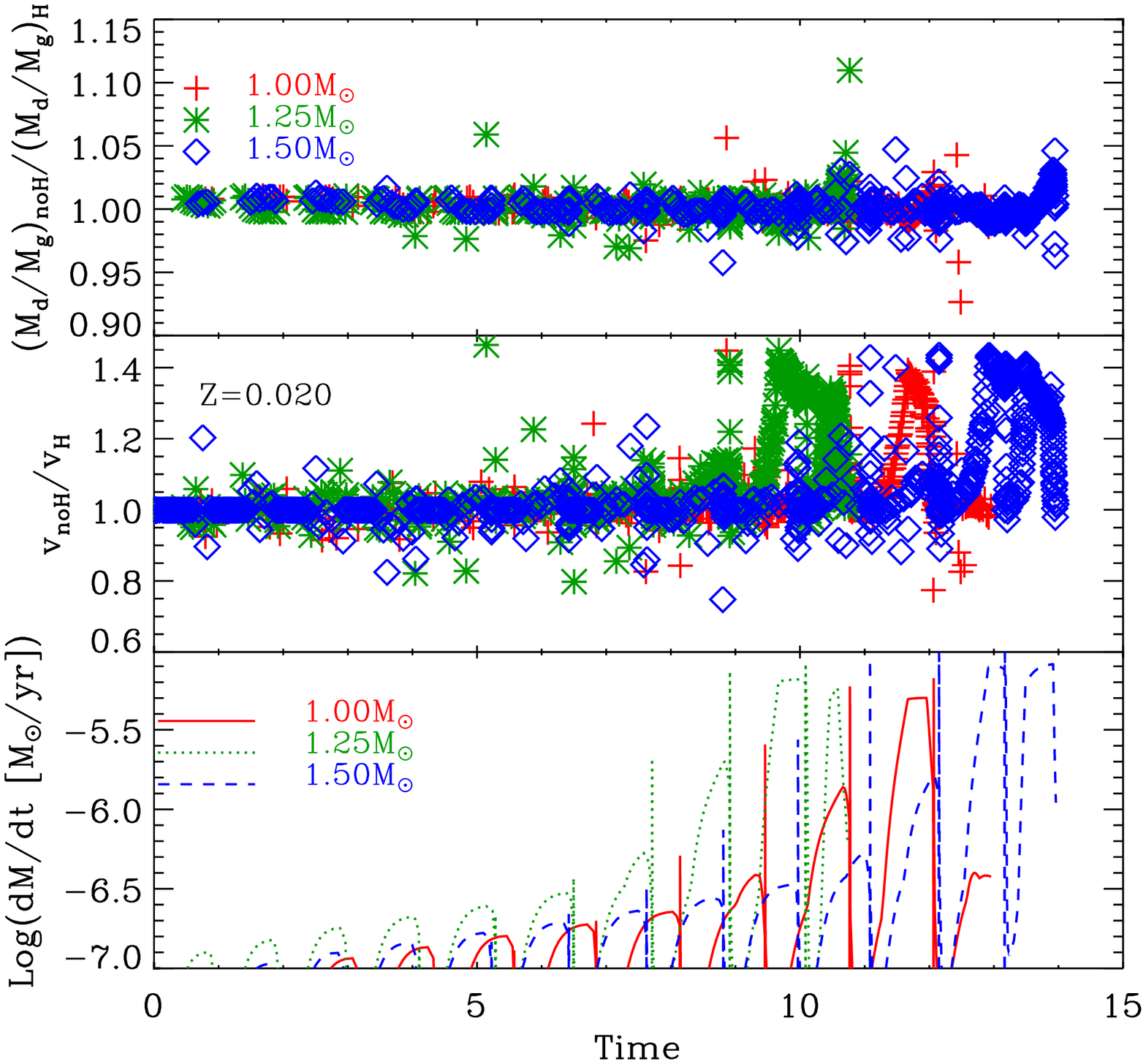}
\caption{\textit{Upper panel}: comparison of dust-to-gas ratios
obtained by neglecting or considering  the heating term of Eq. \ref{energy_balance}.
We show their ratios for the cases with $M=1,\,1.25,\,1.5$~M$_\odot$ and metallicity $Z=0.02$.
\textit{Middle panel}: the same as in the upper panel but for the expansion velocities.
\textit{Lower panel}: mass-loss rate as a function of time. This quantity does not depend on the specific model assumption,
but is characteristic of the TP-AGB phase considered.}
\label{res_noH_H_Z0.02}
\end{figure}

Since the boundary condition on $\tau$ (Eq.~\ref{taufin}) is at infinity, but
all the other conditions are at the condensation radius, we
solve the system by means of the shooting method thoroughly described in \citet{Nanni13}.

\begin{figure*}
\centering
\includegraphics[width=0.45\textwidth]{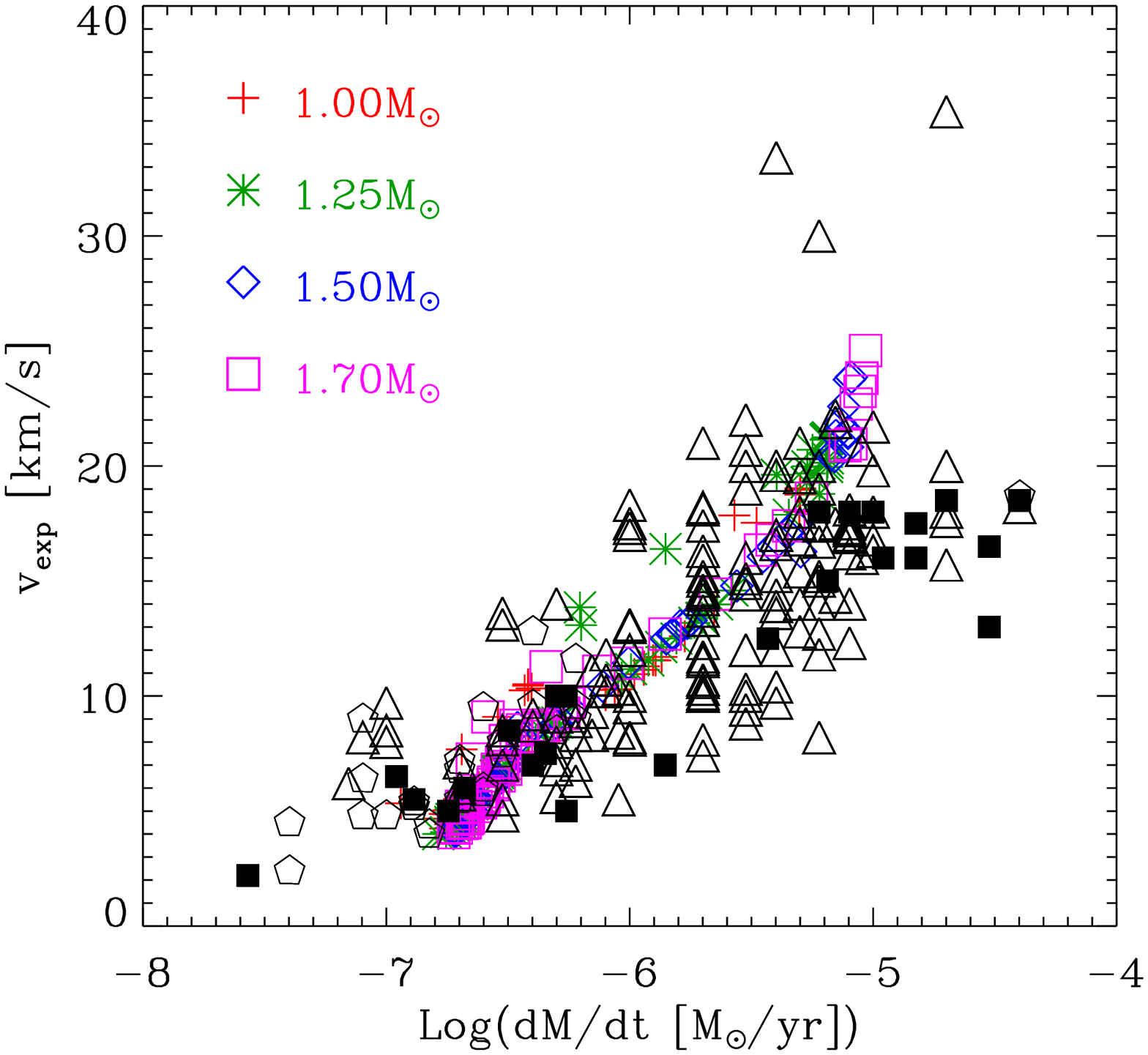}
\includegraphics[width=0.45\textwidth]{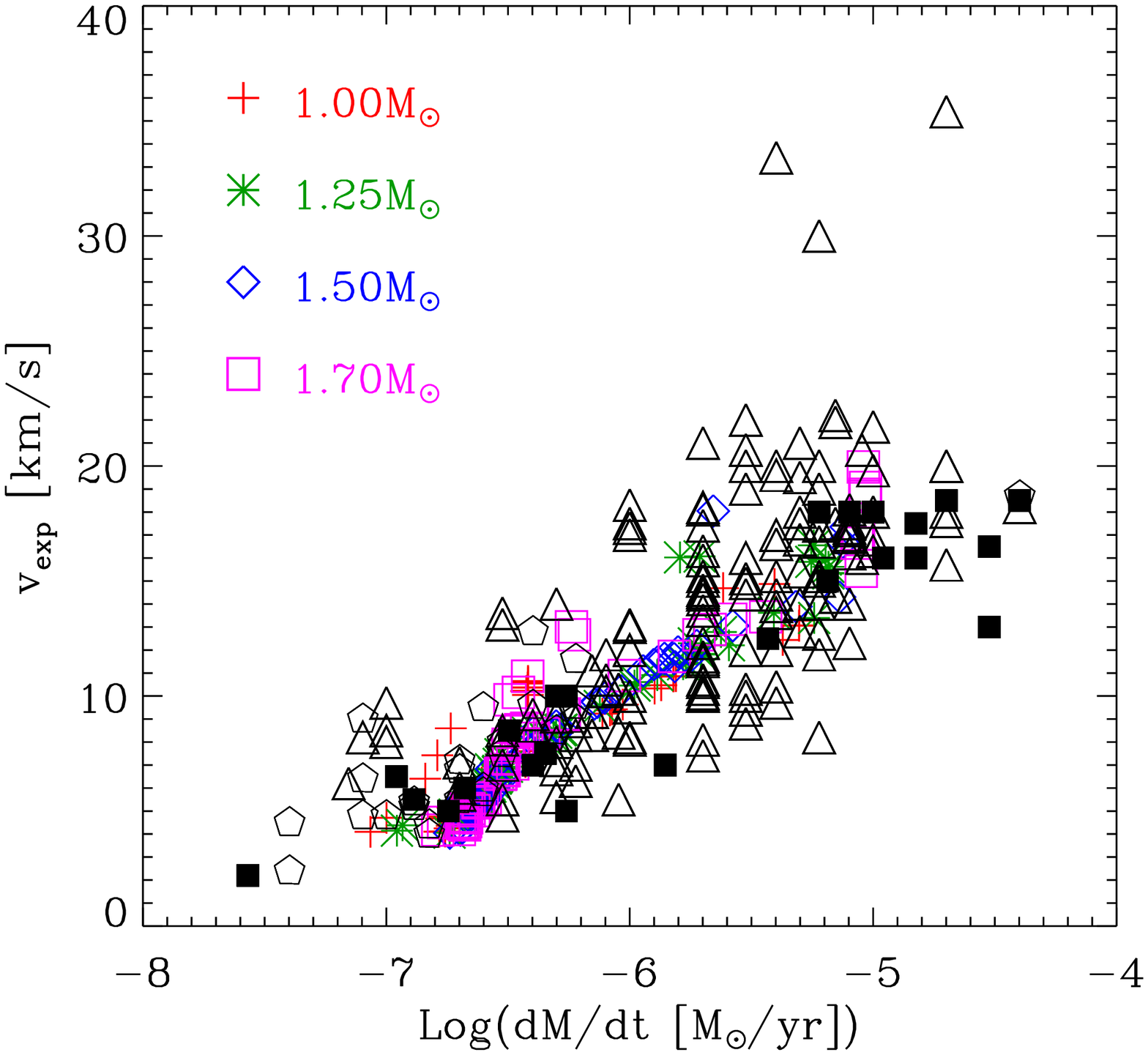}
\caption{Expansion velocities of circumstellar outflows against mass-loss rates
of variable M-stars. Observations of Galactic M-stars by \citet{Loup93} (black triangles),
\citet{Gonzalez03} (black pentagons) and \citet{Schoier13} (black squares)
are compared with predicted expansion velocities for a few selected TP-AGB tracks with $Z=0.02$
for the values of initial stellar masses listed in upper left of each figure.
\textit{Left panel}: comparison with simulations that do not include the heating effect due to H$_2$ collisions.
\textit{Right panel}: comparison with models that include the effect of the heating.
For this metallicity the results do not change significantly.}
\label{v_dotm_02}
\end{figure*}
The reactions assumed for dust formation are taken from \citet{GS98}, for corundum, and FG06 for all the other dust compounds.
The thermodynamical data for the decomposition rates are taken from \citet{Sharp90}
with the exception of FeSiO$_3$ and SiC for which we use the data taken by \citet{Barin95}.
The values of the sticking coefficients are taken from
laboratory measurements, when
available, or from theoretical computations.
For olivine we adopt $\alpha_{\rm ol}\sim$ $0.2$ determined from
evaporation experiments of forsterite by \citep{Nanni13}.
For iron and quartz the experimental values are $\alpha_{\rm Fe}\sim1$ and
$\alpha_{\rm qu}\sim0.01$, respectively \citep{Landolt68}.
For the growth of SiC a value of $\alpha_{\rm SiC}\sim 1$ has
been determined by \citet{Raback99}.
For pyroxene  we adopt the same value as that used for olivine.
For corundum we chose the maximum possible value, i.e. $\alpha_{\rm co}=1$.
Finally, for amorphous carbon the usual assumption is to adopt $\alpha_{\rm C}=1$.
The effects of adopting a lower value for carbon sticking coefficient
have been discussed in \citet{Nanni13}.

\begin{figure*}
\centering
\includegraphics[width=0.45\textwidth]{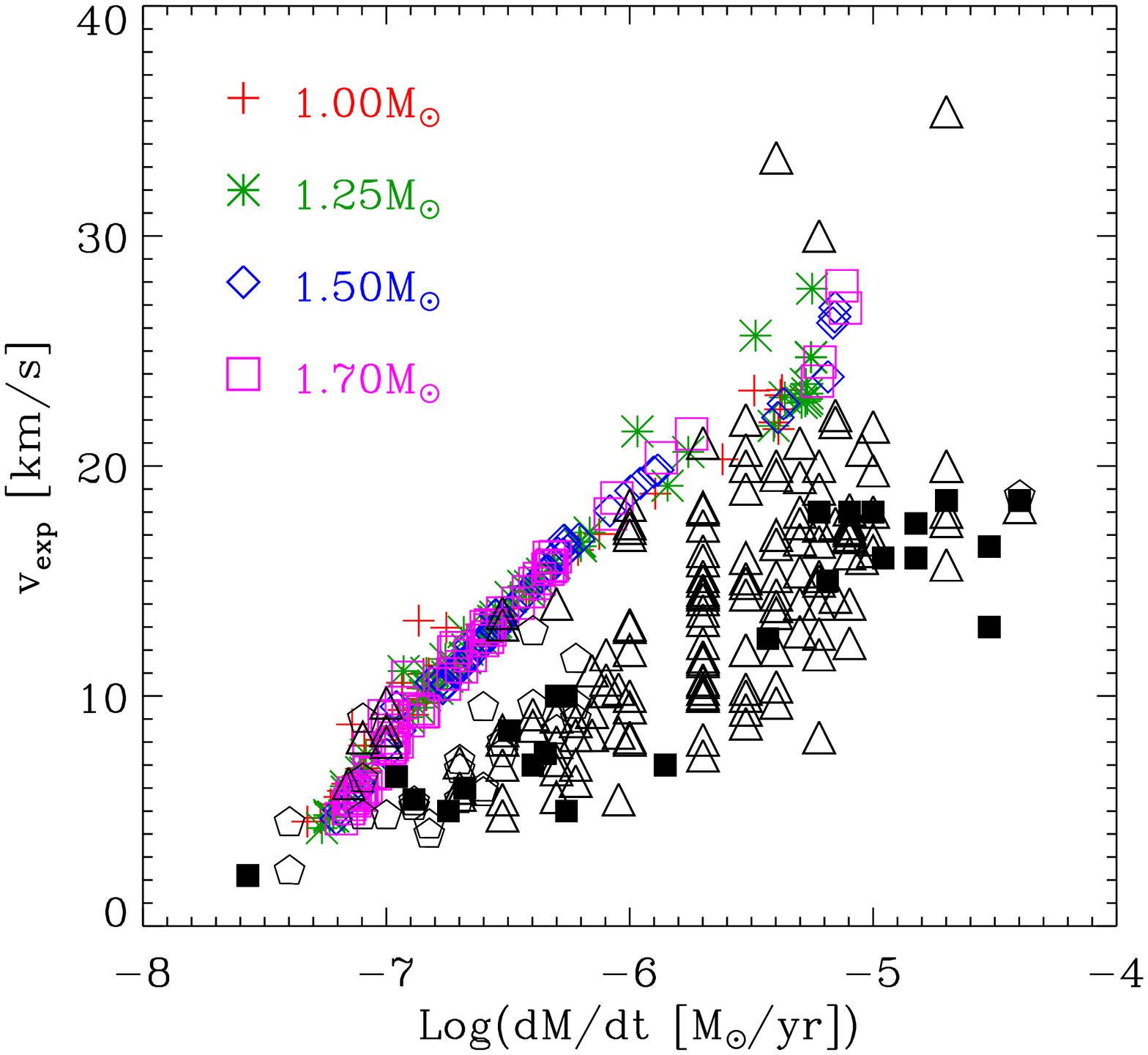}
\includegraphics[width=0.45\textwidth]{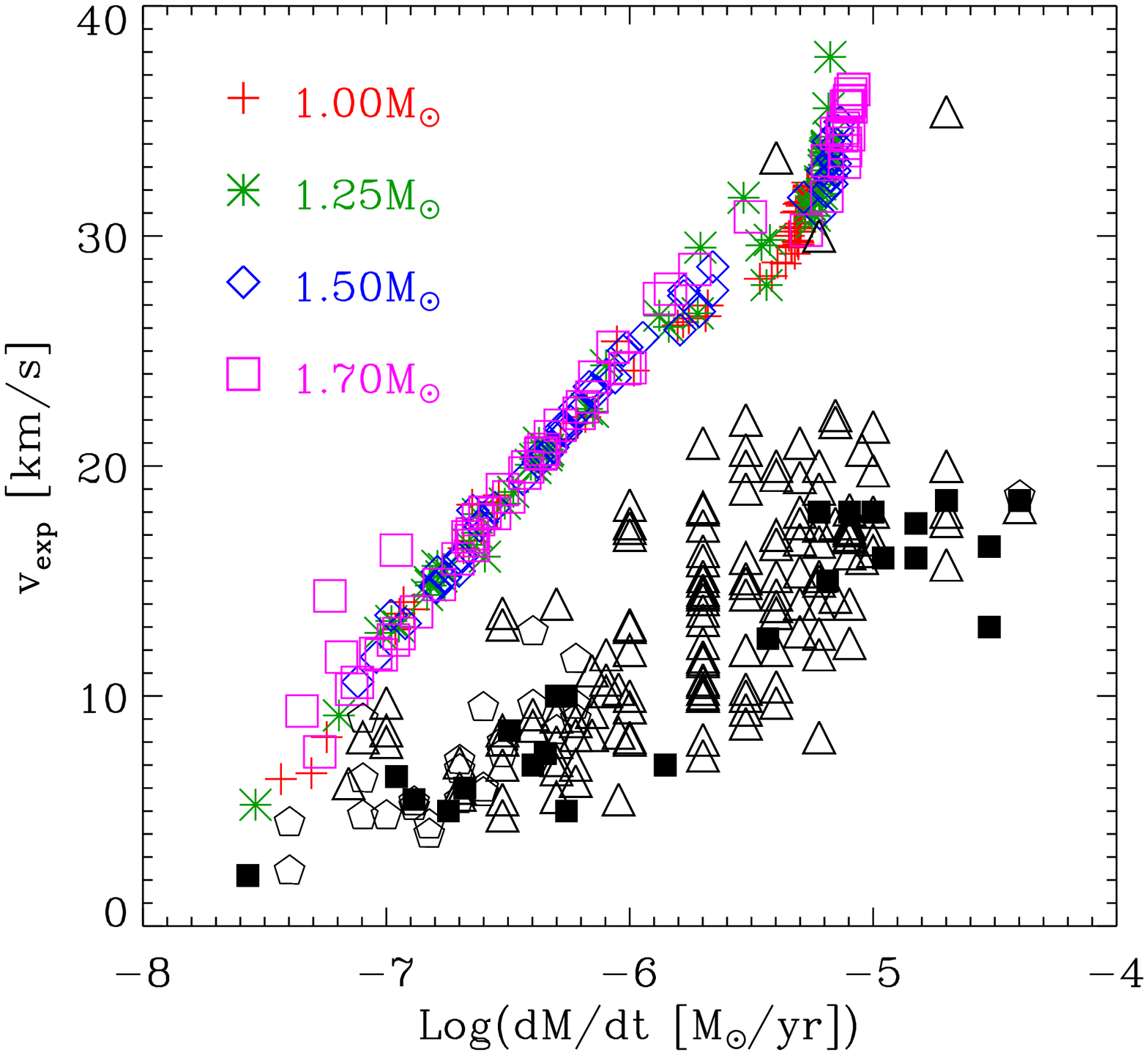}
\caption{The same as the right panel of Fig.~\ref{v_dotm_02} for $Z=0.04$ (left panel) and $Z=0.06$ (right panel).
}
\label{v_dotm_06}
\end{figure*}

For the computation of the opacities
we use a flat dust size distribution
 ($x_g$=0.1) from $a_{\rm min}=0.005$~$\mu$m to $a_{\rm max}=0.18$~$\mu$m.
This choice is consistent  with the grain size distribution obtained from the integration.
The number of initial seeds is set to $\epsilon_s=10^{-13}$ and their initial size is assumed to be $a_0=0.001$~$\mu$m (FG06).
The value of initial velocity of the outflow was arbitrary chosen to be 0.5~km~s$^{-1}$.
If $\Gamma$ never exceeds the unity,
dust is not able to drive the wind --inefficient dust-driven case according to FG06, \citet{Hofner09}--
and we neglect the fraction of dust condensed in these phases.
This approximation does not affect the ejecta of the major dust species
because the mass-loss rates corresponding to inefficient dust-driven wind cases are low.
We also exclude from the calculations the models with mass-loss rates
below 10$^{-8}$~M$_\odot$~yr$^{-1}$ since in these cases,
dust formation is negligible (FG06).

\section{Results}
\label{sec_res}
We follow dust growth in the CSEs of a few selected evolutionary TP-AGB tracks,
extracted from the set of \citet{marigoetal13}.
The differences that arise in the CSEs structure as a consequence of including or neglecting the collisional heating
term, Eq.~(\ref{collision_rate}), in Eq.~(\ref{energy_balance}),
are shown in Fig.~\ref{CSE_Z0.02} for $Z=0.02$ and in Fig.~\ref{CSE_Z0.06}  for $Z=0.06$, respectively.
In the case of $Z=0.02$ we select two cases, one with an intermediate
mass-loss rate ($\sim 10^{-6}$~M$_\odot$~yr$^{-1}$) and the other with a high mass-loss rate ($\sim 10^{-5}$~M$_\odot$~yr$^{-1}$).
In the panels of Fig.~\ref{CSE_Z0.02}, we plot the run of the gas and dust temperature,
the condensation fraction of silicates, the expansion velocity and the density,
as a function of the radius within the CSE. The upper panels show the case in which the collisional heating term is neglected
while in the lower ones, we include this contribution to the energy balance equation.
When the heating term is included,  the dust equilibrium temperature at a given radius rises, the sublimation rate increases
(Eq.~\ref{dust_subl}) and, as a consequence, the
condensation point shifts outwards, where this heating term becomes  less important.
As shown in the left panel of
Fig.~\ref{CSE_Z0.02}, the structure of the CSE remains practically unchanged
for a mass loss rate of $\dot{M}\sim 10^{-6}$~M$_\odot$~yr$^{-1}$.
When the heating term is included in the cases representative of a large mass-loss rate, $\dot{M}\sim 10^{-5}$~M$_\odot$~yr$^{-1}$,
the condensation radius increases by about 50\% and the expansion velocity
decreases by 20\%. The final condensation fraction remains almost unchanged.
Finally, we notice that this term does not affect significantly the dust condensation temperature,
which we obtain through the energy balance, but
only the radius at which this condensation temperature is reached.
An overall comparison of the differences in the dust-to-gas ratios and expansion velocities
between the no-heating and heating models is shown in Fig. \ref{res_noH_H_Z0.02} for $M=1,\,1.25,\,1.5$~M$_\odot$~yr$^{-1}$ at $Z=0.02$.
We first notice that the variation of these two quantities in the two schemes is modulated by the mass-loss rate that is
independent of the scheme adopted.
The difference between the heating and no-heating cases is within 5\% for the dust-to-gas ratios, while
for the expansion velocities, the differences are up to 40\% at the highest mass-loss rates.
However, on average, the effects of neglecting the heating term are small. In particular,
the differences between the dust ejecta are almost negligible.

At super-solar metallicity, $Z=0.06$, we should consider that
(i) the gas is characterized by a higher
number density of metals and, (ii) the effective temperatures of the  models
are generally lower than in the solar metallicity case.
If the collisional heating term is neglected,
both circumstances allow dust to
cool down efficiently even at small distances from the photosphere, $r\sim 1.5$~$R_*$.
Here the density and the temperature are higher than in the more external regions,
giving rise to larger optical depths at the condensation point.
In the case computed without the collisional heating term, shown in the left panel of  Fig.~\ref{CSE_Z0.06},
the solution requires an optical depth at the condensation point that is larger than unity,
which is inconsistent with the adopted
gray atmosphere temperature structure (i.e. the gas temperature below the condensation point is
even larger than the effective temperature).
Thus, solving the system of equations without the heating term
at high mass-loss rates for the high metallicity case, $Z=0.06$,
leads to inconsistent results.
The importance of the collisional heating term
can be appreciated by looking at the right panel of  Fig.~\ref{CSE_Z0.06}.
This term prevents dust from condensing too early in the envelope and, setting
a natural threshold to the condensation radius, prevents
the occurrence of inconsistent solutions.
\subsection{Expansion velocities}\label{subsec_vel}
In addition to Fig. \ref{res_noH_H_Z0.02}, we show
in Fig.~\ref{v_dotm_02} the expansion velocities of TP-AGB stars with masses $M=1,\, 1.25,\, 1.5,\, 1.7$~M$_\odot$
and $Z=0.02$, taken from \citet{Nanni13} as shown in Table~\ref{Table:ejecta}, against the corresponding mass-loss rates both for
the no-heating (left panel) and the heating (right panel) scenarios. For each track we plot
a discrete number of points, selected from
a randomly generated uniform distribution of ages that
samples the entire TP-AGB phase of each star.
This selection method better represents the expected expansion velocity distribution.
The models are compared with the velocities observed
in Galactic M-type AGB stars.
Data of
mass-loss rates and expansion velocities
are taken from \citet{Loup93} (black triangles),
\citet{Gonzalez03} (black pentagons) and \citet{Schoier13} (black squares).
Data from \citet{Loup93} and \citet{Schoier13} are derived from
observations of $^{12}$CO and HCN rotational transitions, whereas
\citet{Gonzalez03} obtained their values from the interpretation of SiO rotational transition
lines. Following the above authors, the uncertainties of the velocity measurements are small,
typically $\leq$10\%, while those of the mass-loss rates can be significantly higher
reaching, in some cases, even an order of magnitude.
For the metallicity we have adopted $Z=0.02$, which is the value derived by
\citep{Smith_Lambert85, Lambert86} for their sample of Galactic M and C giants.
The stellar masses are not observationally determined and we assume that they
are mostly below $M\lesssim 2$~M$_\odot$ because  M-giants of higher masses are more rare in our Galaxy.

From inspection of the models and of Fig.~\ref{v_dotm_02} we see first that,
at solar metallicity,  neglecting the heating term does never lead to inconsistent models.
Second, we see that the effects of collisions become important only at the highest mass loss rates,
where they produce a decrease of the expansion velocities of about 10\%-20\%.
Thus, at solar metallicity, collisional heating is of minor importance,
with respect to other uncertainties. At  the lower metallicities the effect is even less important.

At super-solar metallicities, the combined effects of  higher opacities and lower effective temperatures
favour dust condensation very close to the stellar photosphere and the collisional heating term
in Eq.~\ref{energy_balance} is no more negligible.
Furthermore, at supersolar metallicity,
only the model of $M=5$~M$_\odot$ with $Z=0.04$ is able to reach
a C/O ratio larger than unity,
as shown in Figs.~\ref{fig_agbmodz04} and \ref{fig_agbmodz06}.
As a consequence the C-star phase is practically absent
and we will focus the discussion the M-giant stars.

The expansion velocities of our super-metal rich models are plotted
against the mass-loss rates in Fig.~\ref{v_dotm_06}. We used the same method adopted in Fig.~\ref{v_dotm_02}
and we also put the Galactic data only to check where high metallicity stars
are expected in this plot with respect to the ones at solar metallicity.
At increasing metallicity the velocity at a given mass-loss rate are higher because of the larger gas opacity.
We notice that the velocities of the most metal rich models, $Z=0.06$ are definitely larger
than the observed velocities. However a few observed M-type AGB stars, along the highest envelope of the data,
could be compatible with low mass super metal rich stars with $Z\sim0.04$.

\subsection{Condensation fractions, composition, dust sizes and dust mass loss rates}
\begin{figure*}
\centering
\includegraphics[angle=90,width=0.95\textwidth]{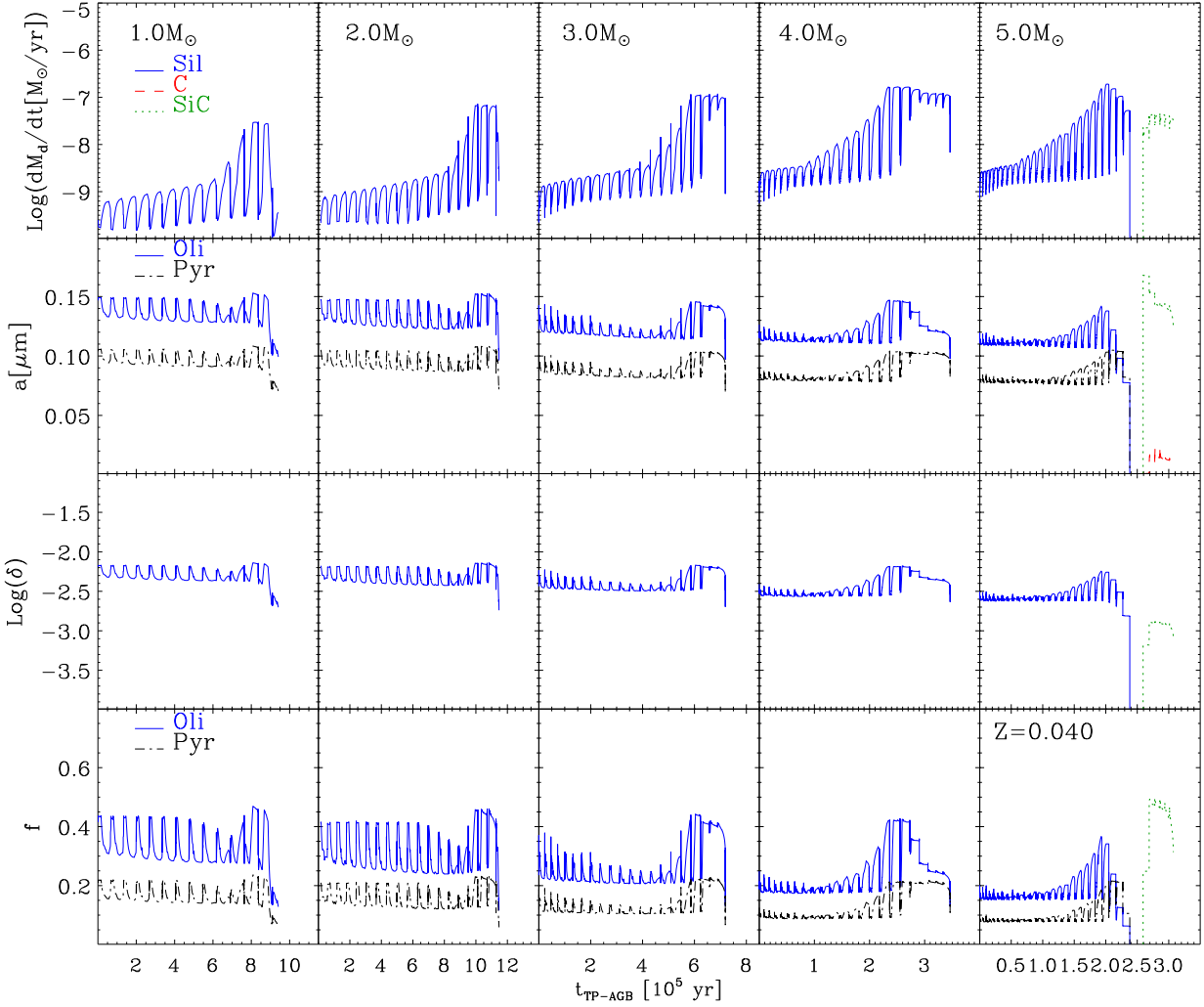}
\caption{Dust properties of selected models of initial metallicity $Z=0.04$,
for various initial masses, as shown in the upper panels.
From top to bottom we depict the dust mass loss rates in M$_\odot$~yr$^{-1}$,
the dust sizes in $\mu$m, the dust-to-gas ratios $\delta$,
and the dust condensation fractions $f$, respectively.
The main dust species are silicates (blue lines). For $M=5$~M$_\odot$
amorphous carbon (red lines) and
SiC (green lines) are also produced.
In some panels silicates are separated into olivine type dust (blue lines)
and pyroxene type dust (black lines) as indicated in the insets. }
\label{fig_agbmod_dustz04}
\end{figure*}

\begin{figure*}
\centering
\includegraphics[angle=90,width=0.95\textwidth]{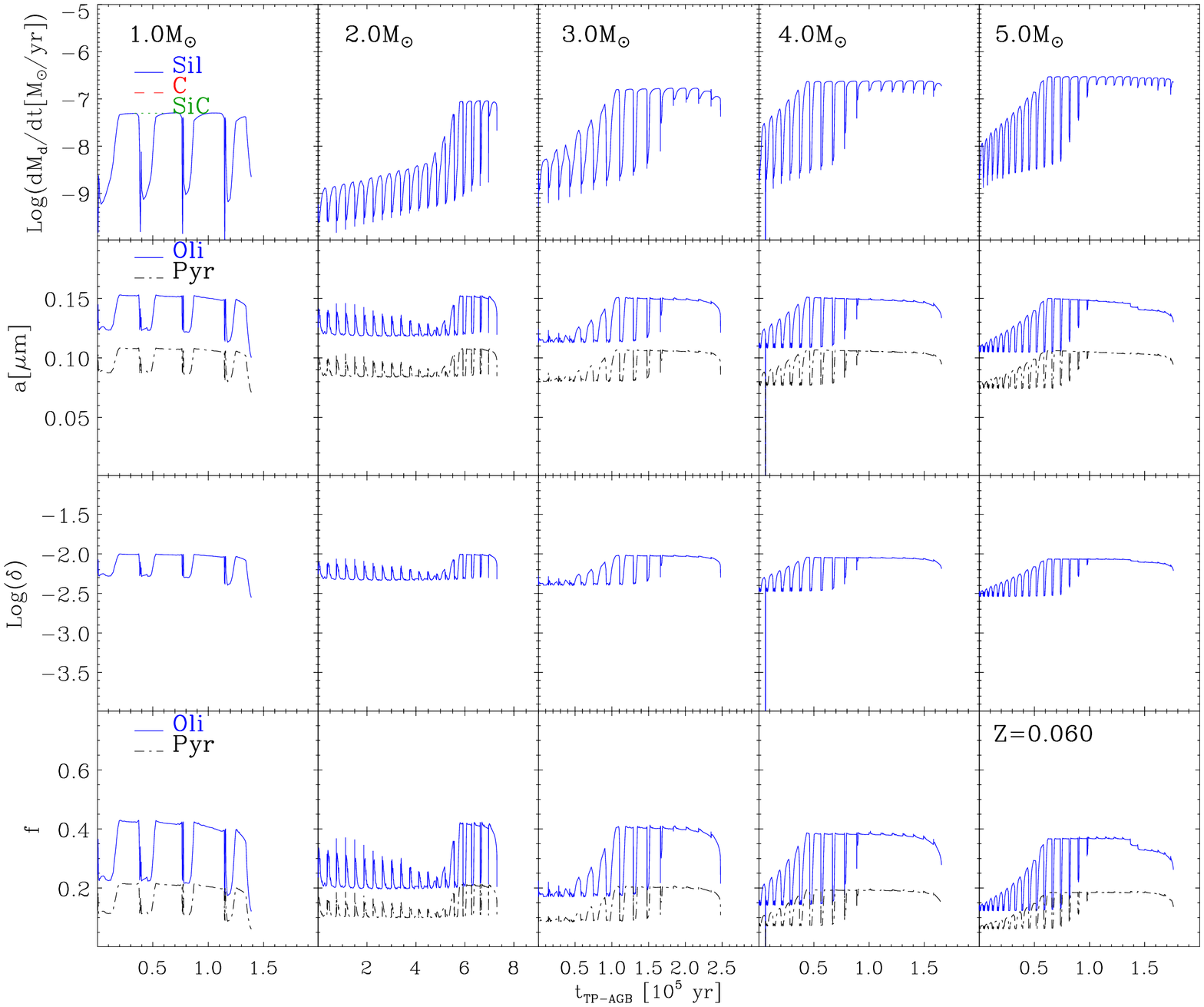}
\caption{The same as in Fig.~\ref{fig_agbmod_dustz04},
but for initial metallicity $Z=0.06$.}
\label{fig_agbmod_dustz06}
\end{figure*}

As already mentioned, super metal rich stars are dominated by silicate dust  production, i.e. olivine and pyroxene.
This is shown in Figs.~\ref{fig_agbmod_dustz04} and \ref{fig_agbmod_dustz06}, for $Z=0.04$ and
$Z=0.06$.
In the same figures, the sizes of dust grains are shown.
The typical value reached is between $a=0.1$~$\mu$m and $a=0.15$~$\mu$m, for all the stellar
masses and metallicities.
These values are similar to those of the solar metallicity case and we confirm that, because of our assumptions,
the maximum value of the size remain almost independent from the metallicity of the star, even up to the most metal rich case
considered here. The reason is that in our model both the initial number of seeds
and the total amount of dust that may condense, scale linearly
with the metallicity (Eq.~\ref{nseeds_M}).

Old super metal rich stars of this kind could populate the nuclear regions of passively evolved
elliptical galaxies \citep{Bertolaetal95} and it is interesting to notice that at a typical age of 10~Gyr,
the turn-off mass is slightly larger than $M=1.1$~M$_\odot$ (see Table~\ref{Table:ejecta}).
These stars can reach
high dust mass loss rates of silicate type, and could thus be responsible of the 9.7~$\mu$m
silicate features discovered with Spitzer in the nuclear regions of
nearby elliptical galaxies \citep{Bressanetal06}.
The observed MIR spectral features are quite broad and could be consistent with a size distribution peaked
toward large grains, as that shown in Figs.~\ref{fig_agbmod_dustz04} and~\ref{fig_agbmod_dustz06}.

The lower two panels in Figs.~\ref{fig_agbmod_dustz04} and~\ref{fig_agbmod_dustz06}
show the dust-to-gas ratios $\delta$, and the condensation fractions $f$, respectively.
Typical values of the silicate dust-to-gas ratios predicted in our models of solar metallicity were around
$\delta\sim$1/200-1/300 while at $Z=0.04$ and $Z=0.06$ we obtain $\delta\leq$1/200 and $\delta\leq$1/100,
respectively. A typical value for the dust-to-gas ratio adopted
 for galactic M-stars is  $\delta\sim$1/200 \citep{Groenewegen_dg98}, and it is usually scaled
linearly with the initial metallicity.
At solar metallicity, our predictions agree fairly well
with the assumed dust-to-gas ratio.
On the other hand, at
super-solar metallicities, the dust-to-gas ratios predicted by our models do not increase linearly with the metallicity.
This can be seen also by noticing that the condensation fractions, plotted in the lowest panels
of the figures, decreases at increasing metallicity.
Taking the $M=1$~M$_\odot$model for comparison, we notice that the condensation fraction
of olivine decreases from $f_{\rm ol}\sim$~$0.45-0.55$ at $Z=0.02$ to  $f_{\rm ol}\sim$ $0.30-0.40$ at $Z=0.04$ and
to $f_{\rm ol}\sim$ $0.20-0.40$ at $Z=0.06$. A similar decrease at increasing $Z$ is shown for the condensation fraction
of the pyroxene. This effect is likely due to the larger outflow velocity, at a given mass-loss rate,
obtained in the more metal rich envelopes, as a consequence of a larger dust opacity.

\section{Dust ejecta}
\label{sec_dust_ejecta}
\begin{figure}
\centering
\includegraphics[width=0.45\textwidth]{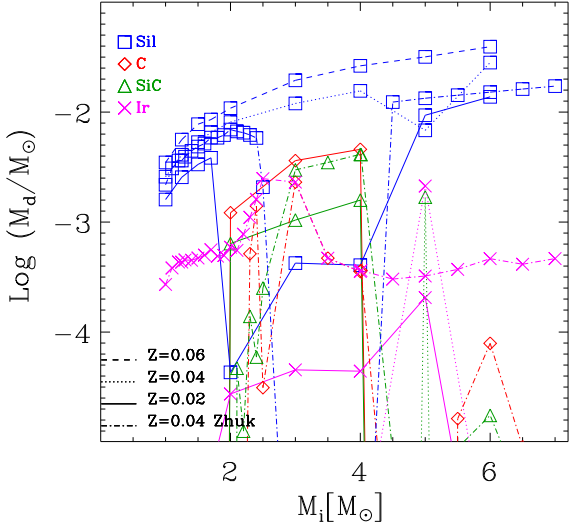}
\caption{Total dust ejecta as a function of
the initial stellar mass and for different initial metallicity,
$Z=0.06$ (dashed lines), $Z=0.04$ (dotted lines) and, for comparison,
$Z=0.02$ from \citet{Nanni13} (solid lines). For the case with
$Z=0.04$ we compare our results with the models by \citet{Zhukovska08}
(dotted-dashed lines.)}
\label{yields}
\end{figure}

The integrated ejecta of the main condensed compounds, i. e.
silicates, amorphous carbon, iron and SiC, along the entire TP-AGB phase
for the different masses and metallicities, are plotted in Fig.~\ref{yields}
and are provided in Table~\ref{Table:ejecta}.
They refer to individual stars and they may not be representative of the
corresponding dust ejecta obtained after convolving with the initial mass function.
We remind that the integration is performed irrespective of the ability of dust to drive or not the stellar wind,
but this unfavorable case occurs in just few models at super-solar metallicity, corresponding to  low
mass-loss rates.
For comparison, we plot in the same Figure also the results of \citet{Nanni13}
for the metallicity $Z=0.02$.
At a given mass, the dust ejecta
increase with the metallicity, albeit not linearly but $\propto Z^{0.7}$, on average.
The silicate ejecta at a given metallicity
increase with the mass, but tend to saturate at high masses
or even decrease for the case of $M=5$~M$_\odot$ at $Z=0.04$.
In the latter case the decrease is due to a C/O ratio that reaches unity toward the end of the evolution
(see Fig.~\ref{fig_agbmod_dustz04}).

At $Z=0.04$, we can compare our results with the ones obtained by \citet{Zhukovska08}.
Below $M=3$~M$_\odot$ and above $M=4$~M$_\odot$,
the resulting dust ejecta are very similar.
On the contrary, in the range between
$M=3$ and $4$~M$_\odot$, the dust production in \citet{Zhukovska08} is dominated
by SiC or by amorphous carbon, rather than by silicates as in our models.
Furthermore, the iron dust ejecta predicted by \citet{Zhukovska08} are always higher than those predicted by our models
and, in some cases they are comparable to
the silicate ones, as for $M=2.5$~M$_\odot$, or to the ones of carbon-bearing dust species,
as in the case of $M=3$~M$_\odot$.
These differences could be due to effects of neglecting chemisputtering  in our models that, anticipating the condensation of silicates,
inhibits the production of iron.
\begin{table*}
\centering
\begin{tabular}{l c c c c c c c c c c c c c}
\hline\hline
$Z$ &\multicolumn{6}{c}{0.04}&\qquad&\multicolumn{6}{c}{0.06}\\
\hline
$M_i$ & Age & Sil&Fe&Al$_2$O$_3$&C&SiC&\qquad&Age & Sil&Fe&Al$_2$O$_3$& C& SiC\\
{[M$_\odot$]} & $\log$~(yr)  &\multicolumn{5}{c}{$\log$~(M/M$_\odot$)}&\qquad& $\log$~(yr)  &\multicolumn{5}{c}{$\log$~(M/M$_\odot$)}\\
\hline
1.00&  10.136 &  $-2.59$&     $-6.41$&     $-4.21$&     -& - & & 10.095 & $-2.46$&   $-7.08$&     $-4.82$&     -& -\\
1.25&  9.793  &  $-2.39$&     $-6.43$&     $-4.13$&     -& - & & 9.765 & $-2.25$&    $ -6.87$&     $-4.62$&     -& -\\
1.50&  9.551 &  $-2.27$&     $-6.26$&     $-4.03$&     -& - & & 9.528  & $-2.11$&     $-6.69$&     $-4.05$&     -& -\\
1.70&  9.392 &  $-2.18$&     $-5.96$&     $-3.94$&     -& - & & 9.372& $-2.07$&     $-6.74$&     $-3.87$&     -& -\\
2.00&  9.226 &  $-2.09$&     $-6.20$&     $-3.84$&     -& - & & 9.188 & $-1.96$&     $-6.63$&     $-3.71$&     -& -\\
3.00&  8.697 &  $-1.92$&     $-6.40$&     $-3.75$&     -& - & & 8.647 & $-1.71$&     $-6.36$&     $-3.89$&     -& -\\
4.00&  8.323 &  $-1.81$&     $-6.05$&     $-3.77$&     -& - & & 8.273  & $-1.58$&    $ -6.23$&     $-4.06$&     -& -\\
5.00&  8.057 &  $-2.17$&     $-2.67$&     $-3.72$&     $-6.06$& $-2.77$ &  & 8.001 & $-1.50$&$-6.09$&     $-3.82$&     -& -\\
6.00&  7.852   & $-1.55$ &       $-6.11$&     $-4.18$&     -& -&  &  7.794    &  $-1.41$&     $-6.00$&     $-4.99$&     -&-\\

\hline\hline
\end{tabular}

\caption{Dust ejecta at $Z=0.04$ and $Z=0.06$.}
\label{Table:ejecta}
\end{table*}
With the new models presented here, we can draw a consistent picture of how dust production
evolves over a wide range of metallicities and for a large mass spectrum.
Particularly useful are the ratios between the total dust ejecta and the gas ejecta,
which, after being convolved with the stellar initial mass function and the star formation rate,
outline the theoretical picture of galactic dust evolution \citep{Dwek07, Valiante09, Dwek11}.
Often, in such models the dust-to-mass ratio is  scaled with the metallicity, i.e.
 $\delta(Z)/\delta_\odot=Z/Z_\odot$, skipping all the
details of the dust production processes.
For $\delta_\odot$ there are different assumptions in literature, but a suitable
average is $\delta_\odot=1/200$. A value $\delta_\odot=Z_\odot$ is sometimes assumed as an extreme case,
if all the metals are supposed to be looked into dust.

\begin{figure}
\centering
\includegraphics[width=0.45\textwidth]{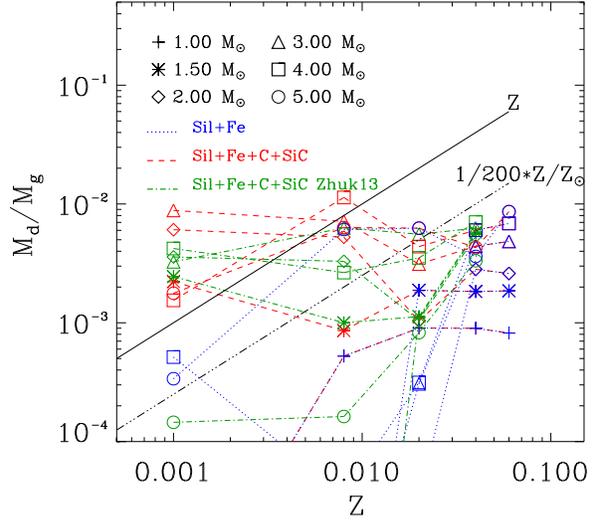}
\caption{Ratios of dust-to-gas ejecta as a function of
the initial stellar mass and initial metallicity. Our results
are compared with the ones obtained by \citet{Zhukovska2013} (Zhukovska, private communication).
}
\label{deltas}
\end{figure}

In Fig.~\ref{deltas} we compare the predictions of the integrated dust-to-gas ratios,
i.e. the ratios between the total dust and gas ejecta $M_{\rm d}/M_{\rm g}$
after integration along the AGB track,
for several masses and as a function of the metallicity.
For the solar and sub-solar metallicities, $Z=0.001,\,0.008,\,0.02$, we
compute the total dust-to-gas from the models presented in \citet{Nanni13}.
We show separately the sum of the silicates and iron ejecta
and the sum of these ejecta plus the carbon and SiC ejecta (the global ejecta), all normalized to the gas ejecta.
This codification allows one to recognize at once the total dust contribution
and the main dust species. Our global ejecta were also compared with the ones obtained by \citet{Zhukovska2013} at $Z=0.001,\, 0.008,\, 0.02,\, 0.04$.
In the same figure, the solid line represents the extreme case $\delta_\odot=Z_\odot$
while, the dotted line, represents the standard assumption $\delta_\odot=1/200$.

A striking feature of this figure is that our global dust-to-gas ejecta
are much less dependent from the metallicity than what usually assumed.
There is of course a modulation with the initial mass but,  considering for
example $M=1.5,\, 2,\, 3$~M$_\odot$, we see that the ratios
$M_{\rm d}/M_{\rm g}$ at super-solar metallicity after an initial decrease at decreasing metallicity,
they rise again
due to the overproduction of carbon at lower metallicity. 
In the models presented by \citet{Zhukovska2013}, a similar trend is found at
$M=1.5,\, 2,\, 4$~M$_\odot$, while, at $M=3$~M$_\odot$,
the dust-to-gas is almost constant in the metallicity range considered.
Our models of $M=1,\, 4,\, 5$~M$_\odot$ show a similar behaviour
down to $Z=0.008$ and, thereafter they show a decline.
In any case the latter models ($M=4,\, 5$~M$_\odot$) at the lowest metallicity,
have ratios that are equal or even slightly larger than the
extreme ratio $\delta_\odot=Z_\odot$.
On the other hand, the dust-to-gas ratios computed by \citet{Zhukovska2013} show a decline for
$M=1$~M$_\odot$  and $5$~M$_\odot$ below $Z=0.02$ and, for this latter mass the total dust-to-gas ratio is well below
the one predicted by our models.
In summary we can say that, at the lowest metallicity
the global dust to gas ejecta may reach and even encompass
the extreme ratio by even an order of magnitude, due to
the efficient carbon dust production while,
at the highest metallicities, the ratios saturate, possibly
due to the abrupt decrease in density produced by the large accelerations
of the dusty driven winds.
This behaviour is similar to the one of the models by \citet{Zhukovska2013}.

\section{Summary and concluding remarks}
\label{sec_discussion}
In this study we perform the analysis of the formation,
evolution, and mineralogy of the dust grains expected to form
in the outflows of TP-AGB stars
with super-solar metallicity, i.e. $Z=0.04$ and $Z=0.06$.
We apply the dust growth formalism presented in \citet{Nanni13},
which is a modification of the one developed by FG06,
to the new TP-AGB evolutionary tracks
of high metallicity $(Z=0.04\,,Z=0.06)$
computed by \citet{marigoetal13}.


\citet{Nanni13} showed that, in order to reproduce
the observed relation between velocities
and mass-loss rates of Galactic M-giant
one has to inhibit  chemisputtering by H$_2$ molecules,
which was instead considered at its full efficiency by FG06 and GS99.
Neglecting this dust destruction process is also in agreement
with recent experimental results of dust evaporation and condensation.
We thus have considered in our formalism only destruction by sublimation.
However, at the high metallicities considered here,
we find necessary to take into account also the coupling between dust and gas particles,
through a collisions heating rate because dust may form in regions
where the temperature and density are so high that
collisions can significantly affect the dust equilibrium temperature.
We check that this effect is not so important in the case of solar and
sub-solar metallicities.

At the high metallicities considered in this work,  $Z=0.04$ and  $Z=0.06$,
the third dredge-up is inefficient and
the TP-AGB lifetimes are short enough to prevent the formation
of carbon stars in most the models here considered.
This is consistent with the extremely low ratio of C- to M-giants
recently found in M31 \citep{Boyeretal13}.
The surface C/O ratio remains below unity
during the whole TP-AGB evolution of all stellar models,
but for the last TP-AGB stages of one massive TP-AGB track.
As a consequence, the dust mineralogy is completely dominated by silicates,
characteristic of M-type stars.
On the other hand, the models by \citet{Zhukovska2013} predict
that, at $Z=0.04$, carbon stars are formed between $M=2.1$~M$_\odot$ and $4.5$~M$_\odot$ and
also around $M=6$~M$_\odot$. However, in their models
dust production is dominated by carbon-bearing species only between $3$~M$_\odot$ and $4$~M$_\odot$.

At super-solar metallicities the expansion velocities are larger than those predicted by
models of solar metallicity. We notice that
some of the Galactic M-giants with the largest
velocities at a given mass-loss rate, could be consistent with
super metal rich stars with $Z=0.04$.

In our models, at a given mass, the dust-to-gas ratios increase with the metallicity,
but with a dependence which is less than linear, likely due to the feedback
produced on the CSE (especially the density) by the large acceleration.
This is also shown by the condensation fractions that, at a given mass, decrease with
increasing metallicity. A similar trend is also predicted by the models by \citet{Zhukovska2013}.

For both metallicities $Z=0.04,\,0.06$,
we present the corresponding dust ejecta, which are mainly under form of silicates.
Finally, by combining them with the total mass lost
by the star during the AGB phase
we obtain the total dust to gas ejecta ratios.
Adding these results to those obtained by \citet{Nanni13},
for the lower metallicities, $Z=0.001,\, 0.008,\, 0.002$,
the following striking global picture emerges.
The total dust to gas ejecta of intermediate mass stars
are much less dependent from the metallicity than what usually assumed.
At the lowest metallicity the total dust to gas ejecta may reach and even encompass
the value of the corresponding initial metallicity by even an order of magnitude, due to
the efficient carbon dust production.
At the highest metallicities, the ratios saturate, possibly
due to the abrupt decrease in density produced by the large accelerations
of the dusty driven winds.
While we do think that the picture at the high metallicities is plausible,
we warn the reader that the issue of the large carbon dust production at the lowest metallicities
is still open and awaits for direct observational confirmations \citep{vanloon08}.

\subsection*{Acknowledgements}
We thank L. Danese, A. Tielens, H. Kobayashi, I. Cherchneff for fruitful discussions,
and   S. Zhukovska for providing her models in electronic form.
We thank the anonymous referee for her/his suggestions.
We acknowledge financial support from contract ASI-INAF I/009/10/0,
and from the Progetto di Ateneo 2012,
CPDA125588/12 funded by the University of Padova.
AB acknowledges financial support from MIUR 2009.

\bibliographystyle{mn2e}
\bibliography{dust}

\begin{thebibliography}{}

\bibitem[\protect\citeauthoryear{Barin \& Platzki}{Barin \&
  Platzki}{1995}]{Barin95}
Barin I.,  Platzki G.,  1995, Thermochemical data of pure substances.
No.~v. 1 in Thermochemical Data of Pure Substances, VCH

\bibitem[\protect\citeauthoryear{{Beelen}, {Cox}, {Benford}, {Dowell},
  {Kov{\'a}cs}, {Bertoldi}, {Omont} \& {Carilli}}{{Beelen}
  et~al.}{2006}]{Beelen06}
{Beelen} A.,  {Cox} P.,  {Benford} D.~J.,  {Dowell} C.~D.,  {Kov{\'a}cs} A.,
  {Bertoldi} F.,  {Omont} A.,    {Carilli} C.~L.,  2006, \apj, 642, 694

\bibitem[\protect\citeauthoryear{{Begemann}, {Dorschner}, {Henning},
  {Mutschke}, {Guertler}, {Koempe} \& {Nass}}{{Begemann}
  et~al.}{1997}]{Begemann97}
{Begemann} B.,  {Dorschner} J.,  {Henning} T.,  {Mutschke} H.,  {Guertler} J.,
  {Koempe} C.,    {Nass} R.,  1997, \apj, 476, 199

\bibitem[\protect\citeauthoryear{{Bell} \& {Lin}}{{Bell} \&
  {Lin}}{1994}]{Bell94}
{Bell} K.~R.,  {Lin} D.~N.~C.,  1994, \apj, 427, 987

\bibitem[\protect\citeauthoryear{{Bertola}, {Bressan}, {Burstein}, {Buson},
  {Chiosi} \& {di Serego Alighieri}}{{Bertola} et~al.}{1995}]{Bertolaetal95}
{Bertola} F.,  {Bressan} A.,  {Burstein} D.,  {Buson} L.~M.,  {Chiosi} C.,
  {di Serego Alighieri} S.,  1995, \apj, 438, 680

\bibitem[\protect\citeauthoryear{{Bertoldi}, {Cox}, {Neri}, {Carilli},
  {Walter}, {Omont}, {Beelen}, {Henkel}, {Fan}, {Strauss} \&
  {Menten}}{{Bertoldi} et~al.}{2003}]{Bertoldi03}
{Bertoldi} F.,  {Cox} P.,  {Neri} R.,  {Carilli} C.~L.,  {Walter} F.,  {Omont}
  A.,  {Beelen} A.,  {Henkel} C.,  {Fan} X.,  {Strauss} M.~A.,    {Menten}
  K.~M.,  2003, \aap, 409, L47

\bibitem[\protect\citeauthoryear{{Bladh}, {H{\"o}fner}, {Nowotny}, {Aringer} \&
  {Eriksson}}{{Bladh} et~al.}{2013}]{Bladh12}
{Bladh} S.,  {H{\"o}fner} S.,  {Nowotny} W.,  {Aringer} B.,    {Eriksson} K.,
  2013, \aap, 553, A20

\bibitem[\protect\citeauthoryear{{Boyer}, {Girardi}, {Marigo}, {Williams},
  {Aringer}, {Nowotny}, {Rosenfield}, {Dorman}, {Guhathakurta}, {Dalcanton},
  {Melbourne}, {Olsen} \& {Weisz}}{{Boyer} et~al.}{2013}]{Boyeretal13}
{Boyer} M.~L.,  {Girardi} L.,  {Marigo} P.,  {Williams} B.~F.,  {Aringer} B.,
  {Nowotny} W.,  {Rosenfield} P.,  {Dorman} C.~E.,  {Guhathakurta} P.,
  {Dalcanton} J.~J.,  {Melbourne} J.~L.,  {Olsen} K.~A.~G.,    {Weisz} D.~R.,
  2013, \apj, 774, 83

\bibitem[\protect\citeauthoryear{{Bressan}, {Granato} \& {Silva}}{{Bressan}
  et~al.}{1998}]{Bressan98}
{Bressan} A.,  {Granato} G.~L.,    {Silva} L.,  1998, \aap, 332, 135

\bibitem[\protect\citeauthoryear{{Bressan}, {Marigo}, {Girardi}, {Salasnich},
  {Dal Cero}, {Rubele} \& {Nanni}}{{Bressan} et~al.}{2012}]{Bressanetal12}
{Bressan} A.,  {Marigo} P.,  {Girardi} L.,  {Salasnich} B.,  {Dal Cero} C.,
  {Rubele} S.,    {Nanni} A.,  2012, \mnras, 427, 127

\bibitem[\protect\citeauthoryear{{Bressan}, {Panuzzo}, {Buson}, {Clemens},
  {Granato}, {Rampazzo}, {Silva}, {Valdes}, {Vega} \& {Danese}}{{Bressan}
  et~al.}{2006}]{Bressanetal06}
{Bressan} A.,  {Panuzzo} P.,  {Buson} L.,  {Clemens} M.,  {Granato} G.~L.,
  {Rampazzo} R.,  {Silva} L.,  {Valdes} J.~R.,  {Vega} O.,    {Danese} L.,
  2006, \apjl, 639, L55

\bibitem[\protect\citeauthoryear{{Cherchneff}, {Barker} \&
  {Tielens}}{{Cherchneff} et~al.}{1992}]{Cherchneff92}
{Cherchneff} I.,  {Barker} J.~R.,    {Tielens} A.~G.~G.~M.,  1992, \apj, 401,
  269

\bibitem[\protect\citeauthoryear{{Clemens}, {Bressan}, {Panuzzo}, {Rampazzo},
  {Silva}, {Buson} \& {Granato}}{{Clemens} et~al.}{2009}]{Clemensetal09}
{Clemens} M.~S.,  {Bressan} A.,  {Panuzzo} P.,  {Rampazzo} R.,  {Silva} L.,
  {Buson} L.,    {Granato} G.~L.,  2009, \mnras, 392, 982

\bibitem[\protect\citeauthoryear{{Clemens}, {Panuzzo}, {Rampazzo}, {Vega} \&
  {Bressan}}{{Clemens} et~al.}{2011}]{Clemensetal11}
{Clemens} M.~S.,  {Panuzzo} P.,  {Rampazzo} R.,  {Vega} O.,    {Bressan} A.,
  2011, \mnras, 412, 2063

\bibitem[\protect\citeauthoryear{{Di Criscienzo}, {Dell'Agli}, {Ventura},
  {Schneider}, {Valiante}, {La Franca}, {Rossi}, {Gallerani} \& {Maiolino}}{{Di
  Criscienzo} et~al.}{2013}]{DiCriscienzo_etal13}
{Di Criscienzo} M.,  {Dell'Agli} F.,  {Ventura} P.,  {Schneider} R.,
  {Valiante} R.,  {La Franca} F.,  {Rossi} C.,  {Gallerani} S.,    {Maiolino}
  R.,  2013, \mnras, 433, 313

\bibitem[\protect\citeauthoryear{{Duschl}, {Gail} \& {Tscharnuter}}{{Duschl}
  et~al.}{1996}]{Duschl96}
{Duschl} W.~J.,  {Gail} H.-P.,    {Tscharnuter} W.~M.,  1996, \aap, 312, 624

\bibitem[\protect\citeauthoryear{{Dwek} \& {Cherchneff}}{{Dwek} \&
  {Cherchneff}}{2011}]{Dwek11}
{Dwek} E.,  {Cherchneff} I.,  2011, \apj, 727, 63

\bibitem[\protect\citeauthoryear{{Dwek}, {Galliano} \& {Jones}}{{Dwek}
  et~al.}{2007}]{Dwek07}
{Dwek} E.,  {Galliano} F.,    {Jones} A.~P.,  2007, Nuovo Cimento B Serie, 122,
  959

\bibitem[\protect\citeauthoryear{{Eales}, {Lilly}, {Webb}, {Dunne}, {Gear},
  {Clements} \& {Yun}}{{Eales} et~al.}{2000}]{Eales00}
{Eales} S.,  {Lilly} S.,  {Webb} T.,  {Dunne} L.,  {Gear} W.,  {Clements} D.,
   {Yun} M.,  2000, \aj, 120, 2244

\bibitem[\protect\citeauthoryear{{Ferrarotti} \& {Gail}}{{Ferrarotti} \&
  {Gail}}{2006}]{FG06}
{Ferrarotti} A.~S.,  {Gail} H.-P.,  2006, \aap, 447, 553

\bibitem[\protect\citeauthoryear{{Gail} \& {Sedlmayr}}{{Gail} \&
  {Sedlmayr}}{1986}]{Gail86}
{Gail} H.-P.,  {Sedlmayr} E.,  1986, \aap, 166, 225

\bibitem[\protect\citeauthoryear{{Gail} \& {Sedlmayr}}{{Gail} \&
  {Sedlmayr}}{1998}]{GS98}
{Gail} H.-P.,  {Sedlmayr} E.,  1998, Faraday Discussions, 109, 303

\bibitem[\protect\citeauthoryear{{Gail} \& {Sedlmayr}}{{Gail} \&
  {Sedlmayr}}{1999}]{GS99}
{Gail} H.-P.,  {Sedlmayr} E.,  1999, \aap, 347, 594

\bibitem[\protect\citeauthoryear{{Gall}, {Andersen} \& {Hjorth}}{{Gall}
  et~al.}{2011}]{Gall11}
{Gall} C.,  {Andersen} A.~C.,    {Hjorth} J.,  2011, \aap, 528, A14

\bibitem[\protect\citeauthoryear{{Gonz{\'a}lez Delgado}, {Olofsson},
  {Kerschbaum}, {Sch{\"o}ier}, {Lindqvist} \& {Groenewegen}}{{Gonz{\'a}lez
  Delgado} et~al.}{2003}]{Gonzalez03}
{Gonz{\'a}lez Delgado} D.,  {Olofsson} H.,  {Kerschbaum} F.,  {Sch{\"o}ier}
  F.~L.,  {Lindqvist} M.,    {Groenewegen} M.~A.~T.,  2003, \aap, 411, 123

\bibitem[\protect\citeauthoryear{{Goumans} \& {Bromley}}{{Goumans} \&
  {Bromley}}{2012}]{Goumans12}
{Goumans} T.~P.~M.,  {Bromley} S.~T.,  2012, \mnras, 420, 3344

\bibitem[\protect\citeauthoryear{{Groenewegen} \& {de Jong}}{{Groenewegen} \&
  {de Jong}}{1998}]{Groenewegen_dg98}
{Groenewegen} M.~A.~T.,  {de Jong} T.,  1998, \aap, 337, 797

\bibitem[\protect\citeauthoryear{{Groenewegen}, {Whitelock}, {Smith} \&
  {Kerschbaum}}{{Groenewegen} et~al.}{1998}]{Groenewegen98}
{Groenewegen} M.~A.~T.,  {Whitelock} P.~A.,  {Smith} C.~H.,    {Kerschbaum} F.,
   1998, \mnras, 293, 18

\bibitem[\protect\citeauthoryear{{Hanner}}{{Hanner}}{1988}]{Hanner88}
{Hanner} M.,  1988, Technical report, {Grain optical properties}

\bibitem[\protect\citeauthoryear{{H{\"o}fner}}{{H{\"o}fner}}{2008}]{Hofner_size08}
{H{\"o}fner} S.,  2008, Physica Scripta Volume T, 133, 014007

\bibitem[\protect\citeauthoryear{{H{\"o}fner}}{{H{\"o}fner}}{2009}]{Hofner09}
{H{\"o}fner} S.,  2009, in {Henning} T.,  {Gr{\"u}n} E.,   {Steinacker} J.,
  eds, Cosmic Dust - Near and Far Vol.~414 of Astronomical Society of the
  Pacific Conference Series.
p.~3

\bibitem[\protect\citeauthoryear{{Jeong}, {Winters}, {Le Bertre} \&
  {Sedlmayr}}{{Jeong} et~al.}{2003}]{Jeong03}
{Jeong} K.~S.,  {Winters} J.~M.,  {Le Bertre} T.,    {Sedlmayr} E.,  2003,
  \aap, 407, 191

\bibitem[\protect\citeauthoryear{{Kimura}, {Mann}, {Biesecker} \&
  {Jessberger}}{{Kimura} et~al.}{2002}]{Kimura02}
{Kimura} H.,  {Mann} I.,  {Biesecker} D.~A.,    {Jessberger} E.~K.,  2002,
  \icarus, 159, 529

\bibitem[\protect\citeauthoryear{{Knapp}}{{Knapp}}{1985}]{Knapp85}
{Knapp} G.~R.,  1985, \apj, 293, 273

\bibitem[\protect\citeauthoryear{{Kobayashi}, {Kimura}, {Watanabe}, {Yamamoto}
  \& {M{\"u}ller}}{{Kobayashi} et~al.}{2011}]{Kobayashi11}
{Kobayashi} H.,  {Kimura} H.,  {Watanabe} S.-I.,  {Yamamoto} T.,
  {M{\"u}ller} S.,  2011, Earth, Planets, and Space, 63, 1067

\bibitem[\protect\citeauthoryear{{Kobayashi}, {Watanabe}, {Kimura} \&
  {Yamamoto}}{{Kobayashi} et~al.}{2009}]{Kobayashi09}
{Kobayashi} H.,  {Watanabe} S.-I.,  {Kimura} H.,    {Yamamoto} T.,  2009,
  \icarus, 201, 395

\bibitem[\protect\citeauthoryear{{Lambert}, {Gustafsson}, {Eriksson} \&
  {Hinkle}}{{Lambert} et~al.}{1986}]{Lambert86}
{Lambert} D.~L.,  {Gustafsson} B.,  {Eriksson} K.,    {Hinkle} K.~H.,  1986,
  \apjs, 62, 373

\bibitem[\protect\citeauthoryear{Landolt-B\"ornstein}{Landolt-B\"ornstein}{1968}]{Landolt68}
Landolt-B\"ornstein 1968, In: Sch\"afer K. (ed.) Zahlenwerte und Funktionen..
No. v. 5b in Zahlenwerte und Funktionen, Springer-Verlag, Heidelberg

\bibitem[\protect\citeauthoryear{{Leksina} \& {Penkina}}{{Leksina} \&
  {Penkina}}{1967}]{Leksina67}
{Leksina} I.,  {Penkina} N.,  1967, Fizik. Metall. Metalloved., 23, 344

\bibitem[\protect\citeauthoryear{{Lilly}, {Eales}, {Gear}, {Hammer}, {Le
  F{\`e}vre}, {Crampton}, {Bond} \& {Dunne}}{{Lilly} et~al.}{1999}]{Lilly99}
{Lilly} S.~J.,  {Eales} S.~A.,  {Gear} W.~K.~P.,  {Hammer} F.,  {Le F{\`e}vre}
  O.,  {Crampton} D.,  {Bond} J.~R.,    {Dunne} L.,  1999, \apj, 518, 641

\bibitem[\protect\citeauthoryear{{Loup}, {Forveille}, {Omont} \& {Paul}}{{Loup}
  et~al.}{1993}]{Loup93}
{Loup} C.,  {Forveille} T.,  {Omont} A.,    {Paul} J.~F.,  1993, \aaps, 99, 291

\bibitem[\protect\citeauthoryear{{Lucy}}{{Lucy}}{1971}]{Lucy71}
{Lucy} L.~B.,  1971, \apj, 163, 95

\bibitem[\protect\citeauthoryear{{Lucy}}{{Lucy}}{1976}]{Lucy76}
{Lucy} L.~B.,  1976, \apj, 205, 482

\bibitem[\protect\citeauthoryear{{Maiolino}, {Schneider}, {Oliva}, {Bianchi},
  {Ferrara}, {Mannucci}, {Pedani} \& {Roca Sogorb}}{{Maiolino}
  et~al.}{2004}]{Maiolino04}
{Maiolino} R.,  {Schneider} R.,  {Oliva} E.,  {Bianchi} S.,  {Ferrara} A.,
  {Mannucci} F.,  {Pedani} M.,    {Roca Sogorb} M.,  2004, \nat, 431, 533

\bibitem[\protect\citeauthoryear{{Marchenko}}{{Marchenko}}{2006}]{Marchenko06}
{Marchenko} S.~V.,  2006, in {Lamers} H.~J.~G.~L.~M.,  {Langer} N.,  {Nugis}
  T.,   {Annuk} K.,  eds, Stellar Evolution at Low Metallicity: Mass Loss,
  Explosions, Cosmology Vol.~353 of Astronomical Society of the Pacific
  Conference Series.
p.~299

\bibitem[\protect\citeauthoryear{{Marigo}}{{Marigo}}{2002}]{Marigo_02}
{Marigo} P.,  2002, A\&A, 387, 507

\bibitem[\protect\citeauthoryear{{Marigo} \& {Aringer}}{{Marigo} \&
  {Aringer}}{2009}]{MarigoAringer_09}
{Marigo} P.,  {Aringer} B.,  2009, A\&A, 508, 1539

\bibitem[\protect\citeauthoryear{{Marigo}, {Bressan}, {Nanni}, {Girardi} \&
  {Pumo}}{{Marigo} et~al.}{2013}]{marigoetal13}
{Marigo} P.,  {Bressan} A.,  {Nanni} A.,  {Girardi} L.,    {Pumo} M.~L.,  2013,
  \mnras, 434, 488

\bibitem[\protect\citeauthoryear{{Marigo} \& {Girardi}}{{Marigo} \&
  {Girardi}}{2007}]{MarigoGirardi_07}
{Marigo} P.,  {Girardi} L.,  2007, A\&A, 469, 239

\bibitem[\protect\citeauthoryear{{Matsuura}, {Barlow}, {Zijlstra}, {Whitelock},
  {Cioni}, {Groenewegen}, {Volk}, {Kemper}, {Kodama}, {Lagadec}, {Meixner},
  {Sloan} \& {Srinivasan}}{{Matsuura} et~al.}{2009}]{Matsuura09}
{Matsuura} M.,  {Barlow} M.~J.,  {Zijlstra} A.~A.,  {Whitelock} P.~A.,  {Cioni}
  M.-R.~L.,  {Groenewegen} M.~A.~T.,  {Volk} K.,  {Kemper} F.,  {Kodama} T.,
  {Lagadec} E.,  {Meixner} M.,  {Sloan} G.~C.,    {Srinivasan} S.,  2009,
  \mnras, 396, 918

\bibitem[\protect\citeauthoryear{{Matsuura}, {Woods} \& {Owen}}{{Matsuura}
  et~al.}{2012}]{Matsuura12}
{Matsuura} M.,  {Woods} P.~M.,    {Owen} P.~J.,  2012, \mnras, p.~434

\bibitem[\protect\citeauthoryear{{Mattsson}}{{Mattsson}}{2011}]{Mattsson11}
{Mattsson} L.,  2011, \mnras, 414, 781

\bibitem[\protect\citeauthoryear{{Mayya}, {Rosa-Gonzalez}, {Santiago-Cortes},
  {Rodriguez-Merino}, {Vega}, {Torres-Papaqui}, {Bressan} \&
  {Carrasco}}{{Mayya} et~al.}{2013}]{Mayyaetal13}
{Mayya} Y.~D.,  {Rosa-Gonzalez} D.,  {Santiago-Cortes} M.,  {Rodriguez-Merino}
  L.~H.,  {Vega} O.,  {Torres-Papaqui} J.~P.,  {Bressan} A.,    {Carrasco} L.,
  2013, ArXiv e-prints

\bibitem[\protect\citeauthoryear{{Micha{\l}owski}, {Watson} \&
  {Hjorth}}{{Micha{\l}owski} et~al.}{2010}]{Michalowski10}
{Micha{\l}owski} M.~J.,  {Watson} D.,    {Hjorth} J.,  2010, \apj, 712, 942

\bibitem[\protect\citeauthoryear{{Nagahara}, {Ogawa}, {Ozawa}, {Tamada},
  {Tachibana} \& {Chiba}}{{Nagahara} et~al.}{2009}]{nagaharaetal09}
{Nagahara} H.,  {Ogawa} R.,  {Ozawa} K.,  {Tamada} S.,  {Tachibana} S.,
  {Chiba} H.,  2009, in {Henning} T.,  {Gr{\"u}n} E.,   {Steinacker} J.,  eds,
  Cosmic Dust - Near and Far Vol.~414 of Astronomical Society of the Pacific
  Conference Series.
p.~403

\bibitem[\protect\citeauthoryear{{Nagahara} \& {Ozawa}}{{Nagahara} \&
  {Ozawa}}{1996}]{Nagahara96}
{Nagahara} H.,  {Ozawa} K.,  1996, \gca, 60, 1445

\bibitem[\protect\citeauthoryear{{Nanni}, {Bressan}, {Marigo} \&
  {Girardi}}{{Nanni} et~al.}{2013}]{Nanni13}
{Nanni} A.,  {Bressan} A.,  {Marigo} P.,    {Girardi} L.,  2013, \mnras, 434,
  2390

\bibitem[\protect\citeauthoryear{{Ossenkopf}, {Henning} \&
  {Mathis}}{{Ossenkopf} et~al.}{1992}]{Ossenkopf92}
{Ossenkopf} V.,  {Henning} T.,    {Mathis} J.~S.,  1992, \aap, 261, 567

\bibitem[\protect\citeauthoryear{{P\`{e}gouri\`{e}}}{{P\`{e}gouri\`{e}}}{1988}]{Pegourie88}
{P\`{e}gouri\`{e}} B.,  1988, \aap, 194, 335

\bibitem[\protect\citeauthoryear{{Pipino}, {Fan}, {Matteucci}, {Calura},
  {Silva}, {Granato} \& {Maiolino}}{{Pipino} et~al.}{2011}]{Pipino11a}
{Pipino} A.,  {Fan} X.~L.,  {Matteucci} F.,  {Calura} F.,  {Silva} L.,
  {Granato} G.,    {Maiolino} R.,  2011, \aap, 525, A61

\bibitem[\protect\citeauthoryear{{Pipino} \& {Matteucci}}{{Pipino} \&
  {Matteucci}}{2011}]{Pipino11b}
{Pipino} A.,  {Matteucci} F.,  2011, \aap, 530, A98

\bibitem[\protect\citeauthoryear{R{\aa}back}{R{\aa}back}{1999}]{Raback99}
R{\aa}back P.,  1999, Modeling of the Sublimation Growth of Silicon Carbide
  Crystals.
CSC research reports: Centre for Scientific Computing, Center for Scientific
  Computing

\bibitem[\protect\citeauthoryear{{Robson}, {Priddey}, {Isaak} \&
  {McMahon}}{{Robson} et~al.}{2004}]{Robson04}
{Robson} I.,  {Priddey} R.~S.,  {Isaak} K.~G.,    {McMahon} R.~G.,  2004,
  \mnras, 351, L29

\bibitem[\protect\citeauthoryear{{Sch{\"o}ier}, {Ramstedt}, {Olofsson},
  {Lindqvist}, {Bieging} \& {Marvel}}{{Sch{\"o}ier} et~al.}{2013}]{Schoier13}
{Sch{\"o}ier} F.~L.,  {Ramstedt} S.,  {Olofsson} H.,  {Lindqvist} M.,
  {Bieging} J.~H.,    {Marvel} K.~B.,  2013, \aap, 550, A78

\bibitem[\protect\citeauthoryear{{Sharp} \& {Huebner}}{{Sharp} \&
  {Huebner}}{1990}]{Sharp90}
{Sharp} C.~M.,  {Huebner} W.~F.,  1990, \apjs, 72, 417

\bibitem[\protect\citeauthoryear{{Smith} \& {Lambert}}{{Smith} \&
  {Lambert}}{1985}]{Smith_Lambert85}
{Smith} V.~V.,  {Lambert} D.~L.,  1985, \apj, 294, 326

\bibitem[\protect\citeauthoryear{{Tielens}, {Waters}, {Molster} \&
  {Justtanont}}{{Tielens} et~al.}{1998}]{Tielens98}
{Tielens} A.~G.~G.~M.,  {Waters} L.~B.~F.~M.,  {Molster} F.~J.,    {Justtanont}
  K.,  1998, \apss, 255, 415

\bibitem[\protect\citeauthoryear{{Valiante}, {Schneider}, {Bianchi} \&
  {Andersen}}{{Valiante} et~al.}{2009}]{Valiante09}
{Valiante} R.,  {Schneider} R.,  {Bianchi} S.,    {Andersen} A.~C.,  2009,
  \mnras, 397, 1661

\bibitem[\protect\citeauthoryear{{Valiante}, {Schneider}, {Salvadori} \&
  {Bianchi}}{{Valiante} et~al.}{2011}]{Valiante11}
{Valiante} R.,  {Schneider} R.,  {Salvadori} S.,    {Bianchi} S.,  2011,
  \mnras, 416, 1916

\bibitem[\protect\citeauthoryear{{van Loon}, {Cohen}, {Oliveira}, {Matsuura},
  {McDonald}, {Sloan}, {Wood} \& {Zijlstra}}{{van Loon}
  et~al.}{2008}]{vanloon08}
{van Loon} J.~T.,  {Cohen} M.,  {Oliveira} J.~M.,  {Matsuura} M.,  {McDonald}
  I.,  {Sloan} G.~C.,  {Wood} P.~R.,    {Zijlstra} A.~A.,  2008, \aap, 487,
  1055

\bibitem[\protect\citeauthoryear{{Vassiliadis} \& {Wood}}{{Vassiliadis} \&
  {Wood}}{1993}]{Vassiliadis93}
{Vassiliadis} E.,  {Wood} P.~R.,  1993, \apj, 413, 641

\bibitem[\protect\citeauthoryear{{Ventura}, {Criscienzo}, {Schneider},
  {Carini}, {Valiante}, {D'Antona}, {Gallerani}, {Maiolino} \&
  {Tornamb{\'e}}}{{Ventura} et~al.}{2012}]{ventura12}
{Ventura} P.,  {Criscienzo} M.~D.,  {Schneider} R.,  {Carini} R.,  {Valiante}
  R.,  {D'Antona} F.,  {Gallerani} S.,  {Maiolino} R.,    {Tornamb{\'e}} A.,
  2012, \mnras, 424, 2345

\bibitem[\protect\citeauthoryear{{Woitke}}{{Woitke}}{2006}]{Woitke06}
{Woitke} P.,  2006, \aap, 460, L9

\bibitem[\protect\citeauthoryear{{Yamasawa}, {Habe}, {Kozasa}, {Nozawa},
  {Hirashita}, {Umeda} \& {Nomoto}}{{Yamasawa} et~al.}{2011}]{Yamasawa11}
{Yamasawa} D.,  {Habe} A.,  {Kozasa} T.,  {Nozawa} T.,  {Hirashita} H.,
  {Umeda} H.,    {Nomoto} K.,  2011, \apj, 735, 44

\bibitem[\protect\citeauthoryear{{Zhukovska}, {Gail} \& {Trieloff}}{{Zhukovska}
  et~al.}{2008}]{Zhukovska08}
{Zhukovska} S.,  {Gail} H.-P.,    {Trieloff} M.,  2008, \aap, 479, 453

\bibitem[\protect\citeauthoryear{{Zhukovska} \& {Henning}}{{Zhukovska} \&
  {Henning}}{2013}]{Zhukovska2013}
{Zhukovska} S.,  {Henning} T.,  2013, \aap, 555, A99

\end{thebibliography}
\label{lastpage}
\end{document}